\documentclass[a4paper,12pt,final]{article}
\usepackage{etex}

\usepackage[sc]{mathpazo}

\usepackage[table]{xcolor}

\usepackage{threeparttable}
\usepackage{standalone}
\usepackage{booktabs, dcolumn}
\usepackage{booktabs}

\usepackage{afterpage}

\usepackage{rotating}
\usepackage{tikz}

\usepackage{latexsym}
\usepackage{amsfonts}
\usepackage{amssymb}
\usepackage{amsthm}
\usepackage{amsmath}
\usepackage[mathscr]{eucal}
\usepackage{multicol}
\usepackage{setspace,graphicx}
\usepackage{pdfpages}
\usepackage{pdflscape}

\usepackage{anyfontsize}
\usepackage{t1enc}

\usepackage{caption}
\usepackage{subcaption}
\DeclareCaptionFont{captiofont}{\fontsize{9pt}{10.8pt}\selectfont}
\usepackage{verbatim}
\usepackage[plainpages=false,
breaklinks=true,
colorlinks=true,
urlcolor=gray,
citecolor=blue,
linkcolor=red,
bookmarks=true,
bookmarksopen=true,
bookmarksopenlevel=1,
pdfstartview=FitH,
pdfview=FitH]{hyperref}

\usepackage{color,hyperref}

\usepackage[round]{natbib}

\usepackage[lmargin=1in, rmargin=1in, tmargin=1in, bmargin=1in, paperwidth=220mm, paperheight=290mm]{geometry}

\usepackage{booktabs}
\usepackage{tabulary}
\usepackage{tabularx}
\usepackage{multirow}

\usepackage{enumerate}
\usepackage{sectsty}
\usepackage{lipsum}
\sectionfont{\centering}

\usepackage{tocloft}
\usepackage{titlesec}
\titleformat{\section}
{\large\bfseries}{\thesection}{0.5em}{}

\renewcommand\thesection{\arabic{section}}

\renewcommand\thesubsection{\thesection.\arabic{subsection}}
\titleformat{\subsection}
{\normalfont\itshape}{\thesubsection}{0.5em}{}

\renewcommand\thesubsubsection{\thesection.\arabic{subsection}.\arabic{subsubsection}}
\titleformat{\subsubsection}{\normalfont\itshape}{\thesubsubsection}{0.5em}{}

\usepackage{setspace}

\titlespacing\section{0pt}{10pt plus 4pt minus 2pt}{5pt plus 2pt minus 2pt}
\titlespacing\subsection{0pt}{10pt plus 4pt minus 2pt}{0pt plus 2pt minus 2pt}
\titlespacing\subsubsection{0pt}{10pt plus 4pt minus 2pt}{0pt plus 2pt minus 2pt}

\providecommand{\keywords}[1]{\textbf{Keywords:}  #1}
\providecommand{\JEL}[1]{\textbf{JEL:}  #1}

\makeatletter
\newcommand*{\myfnsymbolsingle}[1]{%
\ensuremath{%
\ifcase#1
\or 
*%
\or 
\dagger
\or 
\ddagger
\or 
\mathsection
\or 
\mathparagraph
\else 
\@ctrerr
\fi
}%
}
\makeatother

\usepackage{alphalph}
\newalphalph{\myfnsymbolmult}[mult]{\myfnsymbolsingle}{}

\renewcommand*{\thefootnote}{%
\myfnsymbolmult{\value{footnote}}%
}

\makeatletter
\def\@xfootnote[#1]{%
\protected@xdef\@thefnmark{#1}%
\@footnotemark\@footnotetext}
\makeatother

\usepackage{bbding} 		

\usepackage[symbol]{footmisc}

\usepackage{chngcntr}

\usepackage{scrwfile}
\TOCclone[\contentsname]{toc}{atoc}
\newcommand\StartAppendixEntries{}
\AfterTOCHead[toc]{%
\renewcommand\StartAppendixEntries{\value{tocdepth}=-10000\relax}%
}
\AfterTOCHead[atoc]{%
\edef\maintocdepth{\the\value{tocdepth}}%
\value{tocdepth}=-10000\relax%
\renewcommand\StartAppendixEntries{\value{tocdepth}=\maintocdepth\relax}%
}
\newcommand*\appendixwithtoc{%
\cleardoublepage
\appendix
\addtocontents{toc}{\protect\StartAppendixEntries}
\listofatoc
}
\usepackage{blindtext}

\newcommand{\tr}{tr}
\newcommand{\p}{per}

\newenvironment{prop*}
  {\ex}
  {\endex}

\newenvironment{remark*}
  {\ex}
  {\endex}

\newenvironment{definition*}
  {\ex}
  {\endex}

\begin{document}

\setcounter{footnote}{0}

\newpage
\hypersetup{colorlinks,linkcolor=black} 

\linespread{1}

\title{\Large \setcounter{footnote}{0} Risks of heterogeneously persistent higher moments\thanks{We are grateful to Anthony Neuberger, Wolfgang Hardle, Antonio Galvao, Rossen Valkanov, Lukas Vacha, Martin Hronec, and the participants at the CFE 2019 for many useful comments, suggestions, and discussions. We gratefully acknowledge the support from the Czech Science Foundation under the EXPRO GX19-28231X project, support from the Grant Agency of Charles University under project No. 1188119, and from Charles University Research Centre program No. UNCE/HUM/035.}
\vspace{20pt}}

\author{\setcounter{footnote}{0}Jozef Barun\'{i}k\thanks{Institute of Economic Studies, Charles University, Opletalova 26, 110 00, Prague, CR and Institute of Information Theory and Automation, Academy of Sciences of the Czech Republic, Pod Vodarenskou Vezi 4, 18200, Prague, Czech Republic, E-mail: \texttt{barunik@fsv.cuni.cz}.}\\
\and
\and
\setcounter{footnote}{6}Josef Kurka\thanks{Institute of Economic Studies, Charles University, Opletalova 26, 110 00, Prague, CR and Institute of Information Theory and Automation, Academy of Sciences of the Czech Republic, Pod Vodarenskou Vezi 4, 18200, Prague, Czech Republic, E-mail: \texttt{josef.kurka@fsv.cuni.cz}.} \\}

\date{\normalsize\vspace{2em} \hspace{2em} First draft: July 2018 \hspace{4em} This draft: \today }

\maketitle

\linespread{1.0}

\begin{abstract}

\noindent  Using intraday data for the cross-section of individual stocks, we show that both transitory and persistent fluctuations in realized market and average idiosyncratic volatility, skewness and kurtosis are differentially priced in the cross-section of asset returns, implying a heterogeneous persistence structure of different sources of higher moment risks. Specifically, we find that idiosyncratic transitory shocks to volatility as well as idiosyncratic persistent shocks to skewness contain strong commonalities that are relevant to investors.
\vspace{10pt}

\keywords{higher moments, transitory, persistent, cross-section of returns}

\JEL{C14, C22, G11, G12}

\end{abstract}

\renewcommand{\thefootnote}{\arabic{footnote}}
\setcounter{footnote}{0}

\newpage
\hypersetup{colorlinks,linkcolor=red} 
\section{Introduction}
\label{sec:intro}

Higher moments, which capture non-normalities in return distributions, have long been recognised as an important source of risk in pricing securities \citep{fama1965portfolio}. More recent work suggests that additional features of the payoff distribution of individual securities may be relevant to understanding differences in asset returns. For example, \cite{amaya2015,neuberger2019skewness} argue that the time variation of moments is an important aspect that induces changes in the investment opportunity set by changing the expectation of future market returns or by changing the risk-return trade-off.\footnote{See also \cite{kelly2014tail,harvey2000,ang2006}. \cite{amaya2015} show that realized skewness measures computed from intraday return data on individual stocks can be used to sort stocks into portfolios with significantly different excess returns, while \cite{boyer2009} and \cite{conrad2013} show that high idiosyncratic skewness in individual stocks is also correlated with positive returns. \cite{ghysels2016invest} present similar results for emerging market indices.} In addition, the risk premium associated with idiosyncratic and systemic counterparts is documented to affect an investor's pricing kernel unequally \citep{langlois2020measuring}. However, these risks associated with higher moments of return distributions are exclusively modelled as constant across frequencies, which severely constrains risk measurement across horizons \citep{bandi2019spectral}. In contrast to this assumption, recent literature documents both theoretical and empirical evidence \footnote{The importance of horizon-specific investor decisions has been recognised in the literature for decades. Horizon choice significantly affects model results in terms of asset pricing \citep{levhari1977capital}, portfolio selection \citep{tobin1965money} and portfolio performance \citep{levy1972portfolio}. Such findings emphasised the importance of capturing heterogeneous investor preferences across investment horizons. Models incorporating such an assumption began to emerge shortly afterwards \citep{gressis1976multiperiod, lee1990heterogeneous}, but the increased attention to modelling horizon-specific risks is a very recent phenomenon \citep{dew2013asset,neuhierl2021frequency,bandi2019spectral}.} that investor preferences are frequency-specific \citep{dew2013asset,neuhierl2021frequency,bandi2019spectral}. It remains an open question how different sources of risk for an investor are high-frequency (low-frequency) fluctuations of higher moments such as skewness or kurtosis associated with transitory (persistent) risks.

The main objective of this paper is to systematically investigate how transitory and persistent higher moment fluctuations are priced into the cross-section of expected stock returns. Since these moment-based risks are highly time-varying and have both transitory (short-term) and persistent (long-term) components, we aim to identify the role of these components using recent advances in financial econometrics, coupled with newly available high-frequency intraday data that allow accurate measurement of the time-varying higher moments.

Why should an investor have heterogeneously persistent preferences for higher instantaneous risks? Risk varies across investment styles as well as frequencies \citep{bandi2019spectral}, so models that assume constant risk across investment horizons generally fail to describe a number of key features, including the pricing of cross-sections, when confronted with data. While long-run risk models \citep{bansal2004} suggest that persistent components of risk are the ones that matter, empirical evidence is mixed, suggesting that they do not fully capture the dynamics in returns. In contrast, \cite{neuhierl2021frequency} argue that a key feature of an asset pricing model should be the ability to decompose risk into frequency-specific components. Higher moments of the return distribution exhibit strong time dynamics \citep{amaya2015}, implying that they contain important transitory and permanent sources of risk. For example, skewness risk, which is often perceived as a manifestation of tail risk or crash risk, may have both transitory and permanent components that can be well linked to the transitory and permanent shocks in the economy, creating heterogeneously persistent risk. Our work is closely related to \cite{neuberger2019skewness}, who suggest how to compute higher moments of long-horizon returns from daily returns. In contrast, we use a cyclical decomposition of fluctuations from intraday data, which provides a full decomposition of information to any frequency band of interest, and we exploit both transitory and persistent components of the higher moments. We see this decomposition as a natural way to explicitly model heterogeneous investment horizons and fully describe their dynamics.

More generally, returns and risks can be decomposed into elements with different degrees of persistence \citep{adrian2008}.  In their seminal work, \cite{bansal2004} suggest frequency decomposition of consumption and dividend growth processes as a key to explaining various puzzles in asset markets. Shocks to consumption at different frequencies have different implications for model outcomes; they enter the pricing kernel with different weights \citep{dew2013asset}, have different effects on asset returns \citep{ortu2013, yu2012using} and lifetime utility \citep{bidder2016long}, and the exposure of firms' cash flows to shocks of different persistence varies \citep{li2016short}. \cite{bandi2019spectral} decompose the betas in the consumption CAPM model, thus disentangling the effect of exposure to market risk for different horizons. \cite{kamara2016horizon} identify the sources of transitory and persistent risks in five cornerstone factors (MKT, SMB, HML, MOM, LIQ) by observing their power to explain the cross-section of expected returns over different horizons.

Investors' different preferences over different horizons justify a certain degree of horizon dependence in their attitude to risk. Several theoretical concepts explain such investor behaviour. For example, myopic loss aversion links an individual's willingness to participate in an investment (alternatively, in a bet, game, etc.) with the valuation horizon \citep[for details see][]{benartzi1995} and thus perceives investor decision making as horizon-specific. Commonly used preferences, e.g. Epstein-Zin \citep{epstein2013substitution}, are described by a discount factor and a risk aversion parameter. Under horizon-dependent risk aversion, the representation of investor preferences needs to be adjusted by adding a patience coefficient \citep{gonzalo2016}. To prevent the results of our model from being driven by the choice of a specific utility function \citep{dittmar2002}, we approximate the stochastic discount factor using the model-free approach \citep[e.g.,][]{dittmar2002, chabi2012pricing}. The empirical model we propose disentangles the short- and long-term characteristics of investors' risk attitudes associated with different sources of transitory and persistent risk by empirically decomposing higher moment risks into different horizons.

Assuming that idiosyncratic risk can be fully diversified, the literature has long assumed that only market risk enters investors' decision making. However, it has been documented both theoretically and empirically \citep[e.g., ][]{amaya2015, jondeau2019average} that idiosyncratic risk is also priced into asset returns. The reason for this may well be that idiosyncratic risk can only be diversified away in an unconnected system of stocks. As argued by \cite{elliott2014financial,barunik2020dynamic}, investors require risk premia for idiosyncratic risk when stocks form a connected network. Another important reason for the relevance of idiosyncratic risk comes from the deliberate under-diversification of investors who, for example, do not hold fully diversified portfolios because they want to take advantage of the extreme positive returns from  positively skewed assets. These investors, in turn, are exposed to average idiosyncratic risk due to the network structure of stock markets and their documented behaviour, which violates rational decision-making in the traditional sense.

While considerable research has examined the time-series relationship between idiosyncratic moments and the cross-section of returns, less attention has been paid to how aggregate moments affect the cross-section of expected returns. The literature documents that while idiosyncratic skewness risk is important, systematic skewness of returns also provides defensive returns in bad times \citep{langlois2020measuring}, and average market skewness is priced into the cross-section of returns \citep{jondeau2019average}. Our work is related to this recent debate, and we further explore how the two sources of risk are priced by investors with heterogeneously persistent preferences.


Our main contribution is to document how the higher moments are priced by investors with heterogeneous investment horizons. We obtain our main empirical results for a cross-section of US firms, using returns on all stocks in the Center for Research in Securities Prices (CRSP) database. The departures from normality in the asset return distribution are most pronounced in the smallest market capitalisation deciles \citep{harvey2000}, so we must not limit our attention to large-cap firms in order to fully assess how the heterogeneously persistent higher moments are priced in the cross-section of stocks. The sample, collected over the period January 2000 to December 2022, uses high frequency data to compute realized moments using one-minute sample prices filtered to 5-minute prices, and we build a database of daily returns and moment factors to test the conditional asset pricing models. We find that both market and idiosyncratic higher moment risks are priced into the cross-section of asset returns with heterogeneous persistence.

\section{Persistence of the Higher Order Moments Risks}
\label{sec:motivation}

The empirical search for explanation of why different assets earn different average returns centers around risk factor models arising from the Euler Equation. Whereas literature documents large number of factors, their overall poor performance \citep{mclean2016, harvey2016} supports the focus on risk factors capturing the properties of asset returns such as moments of distribution. At the same time, researchers document puzzling results when studying volatility, skewness and kurtosis as proxies for risk. We believe that one of the important reasons is that researchers assume homogeneous preferences of investors over different investment horizons. In contrast, we believe investors price these risks with heterogeneous persistence. Below, we provide theoretical discussion connecting persistence in higher moments risks with the cross-section of asset returns and we show how to extract the transitory and persistent components of these risks from high-frequency data.

\subsection{Higher Moments Risks in Asset Pricing}
\label{subsec:hmr}

Volatility is widely perceived as the main indicator of risk on the financial markets and is incorporated in many asset pricing models while higher moments of returns distribution, specifically skewness and kurtosis add crucial information about risk contained in the tails of asset returns distribution. Importance of the tails of distribution in asset pricing is soundly based in the literature and it is manifested mainly by the fact that downside risk plays a pivotal role in predicting returns of multiple asset classes \citep{lettau2014conditional, farago2018downside}.

Normality of the data is a convenient and widely used assumption while higher moments are highly informative once data depart from normality. Non-normalities of return distribution have been recognized in the literature empirically \citep{fama1976,fama1996,bakshi2003stock}, hence the tails of distribution expressed by the higher moments may be able to explain problems documented by standard asset pricing and portfolio selection models like equity premium puzzle\footnote{\cite{mehra1985equity} noted that class of general equilibrium models is not able to explain large average equity risk premia and low risk-free rate observed on US markets.}, deliberate underdiversification\footnote{Investors deliberately hold insufficiently diversified portfolios, although they would be capable of obtaining a sufficient number of assets to fully diversify away the idiosyncratic risk. One of the possible explanations is the desire of investors to hold the ``lottery-like'' assets offering possible extreme positive returns which results in fear of foregoing the opportunity to exploit these returns by becoming completely diversified \citep{simkowitz1978, mitton2007}.} or failures of CAPM\footnote{\cite{harvey2000} note that failures of CAPM are most significant for assets in the lowest deciles of market-cap, i.e. the most significantly skewed assets.} \citep{harvey2000}. More generally, investors deciding under risk often depart from the expected utility framework as they reveal preferences over positively skewed assets \citep{barberis2008}, therefore the preferences of investors are better modelled by the cumulative prospect theory that gives higher weight to the tails of the distribution.\footnote{\cite{barberis2008} claim their model could explain e.g. poor performance of IPOs or success of momentum strategies.}

 The literature documents that higher moments play a significant role in the process of determining the stocks prices. The three-moment CAPM \citep{kraus1976} analytically links the skewness preference with the expected asset returns, \cite{dittmar2002} shows that a pricing kernel incorporating investors preferences over both skewness and kurtosis is necessary to eliminate counterintuitive risk taking, while \cite{chabi2012pricing} derives a pricing kernel containing stochastic volatility, skewness and kurtosis risk. The theoretical concepts linking skewness and/or kurtosis to asset returns also find support in the data. There is evidence that skewness and kurtosis are priced in the financial cross-sections \citep{kraus1976, agarwal2009, chang2013} along with volatility, yet the overall results present various puzzles and we still fail to understand the exact mechanisms propagating these relationships. We believe that building the asset pricing models assuming investors with homogeneous investment horizons is among the  prominent reasons for not being able to fully explain the mechanisms underlying the asset pricing process.

 \subsection{Idiosyncratic and Market Moments}

Another important aspect of the discussion is the type of moment based risk we use in the analysis. A traditional view in the literature is that idiosyncratic moments risks can be diversified away, and only systematic components of moments should be rewarded \citep{harvey2000}. However, enormous literature emphasizes the ability of idiosyncratic risks to predict subsequent returns. Recently, \cite{jondeau2019average} document that average monthly skewness across firms predicts future market returns, and they argue that systematic market skewness is not the main channel by which investor's preferences for skewness affect future market return. In addition, \cite{langlois2020measuring} documents that systematic and idiosyncratic skewness are connected with different expected returns across stocks. While stocks with higher systematic skewness are appealing because they offer defensive returns during bad times, stocks with positive idiosyncratic skewness attract investors seeking high returns regardless of broad market movements, and are connected to a lottery-like payoff. 

Empirical ability of idiosyncratic skewness risk to predict the cross-section of returns has been recognized across many different measures of skewness, specifically option implied measures \citep{conrad2013}, realized measures computed from high-frequency data \citep{amaya2015}, or idiosyncratic skewness forecasted by a time series model \citep{boyer2009}. These findings indicate that investors are willing to accept low returns and high volatility if they are compensated by positive skewness. Such phenomenon is closely connected to deliberate underdiversification \citep{simkowitz1978, mitton2007} that is driven by ``lotto investors" demanding assets with high upside potential. Moreover, preference of investors over ``lottery-like" assets is connected to strong predictive power of maximum past returns \citep{bali2011}, and it plays a central role in explaining the idiosyncratic volatility puzzle \citep{hou2016have}.\footnote{Idiosyncratic volatility puzzle is a phenomenon observed by \cite{ang2006}, who document a negative relationship between idiosyncratic volatility and returns. This is very puzzling as investors should require positive risk premia, if any, for idiosyncratic volatility. However, high idiosyncratic volatility indicates possible high future exposure to idiosyncratic skewness \citep{boyer2009}. Preference for right skewed assets along with market frictions holds a prominent place amongst explanations of idiosyncratic volatility puzzle \citep{hou2016have}.}

Generally, there is evidence that idiosyncratic higher moments can help in explaining multiple financial market puzzles. We contribute to this debate by assessing the role of both market and average idiosyncratic moments with respect to their short-term as well as long-term fluctuations. In other words, we investigate how the transitory and persistent fluctuations of average idiosyncratic and market higher moments are priced in the cross-section of stocks.
 
\subsection{Persistence of the Higher Moment Risks}
\label{subsec:persistence}

Higher moments are exclusively assumed as a risk that is constant across investment horizons in the literature. This is too restrictive for the data, instead the risk factors should be modeled as having heterogeneous impact across investment cycles. Our main aim is to investigate how the transitory as well as persistent fluctuations of higher moments matter in the cross-section of returns. This endeavor stems from the recent discussion which points to changing risk attitudes across investment styles as well as frequencies \citep{bandi2019spectral,neuhierl2021frequency} and suggests that a key feature of an asset pricing model should be the ability to decompose risk into frequency-specific components with different persistence. 

Decades of research have been devoted to the ability of higher moments risks to price the cross-section of equity returns, yet the overall results are still puzzling especially regarding the role of volatility and kurtosis. As discussed above, risks should not be aggregated across investment horizons due to the heterogeneity of risk attitudes connected to different characteristics of investors as well as other aspects of investor preferences. We propose to allow the heterogeneity of investors in terms of investment horizons and assess the connection between the fluctuations in higher moments risks and asset returns based on the persistence of these fluctuations.

To our knowledge, the ability of the transitory and persistent fluctuations in higher moments to price stocks returns has not been evaluated. Similarly to the notion of spectral factor models \citep{bandi2019spectral} that decompose CAPM beta into several frequencies, we decompose the higher moments to heterogeneously persistent components. Relaxing the assumptions of homogeneous investment horizons and uncovering the persistence underlying the higher moments risks should be a crucial step towards explaining the ongoing puzzles connected to the performance of higher moments risk factors.

We are looking at how transitory and persistent shocks to the different higher moments are priced in the cross-section of stocks. We are able to identify the persistence of shocks to the particular moment that plays a role in determining the returns through the pricing kernel. In the next sections we explain how we can naturally model heterogeneous investment horizons and persistence in the higher moments risks using frequency decomposition. We also show the connection between the measures of transitory and persistent higher moments risks, and high-frequency financial data.

\subsection{Modelling persistence}
\label{subsec:decomposition}

Assume that a higher moment $MMT_t$ has two orthogonal components capturing economic cycles shorter than $2^j$ periods and longer than $2^j$ periods (for example months) for $j \ge 1$. These represent the short-term and long-term components capturing transitory and permanent information respectively. \cite{bandi2019spectral} show formally that it is always possible to decompose covariance-stationary time-series in such a way that these two components are orthogonal, they are non anticipative, and hence suitable for out of sample applications. These are key for the purpose of using such factors in asset pricing models.

Assuming the higher moment is a covariance-stationary time-series, we can decompose a higher moment risk factor into transitory and persistent components as
\begin{eqnarray}
 MMT_t &=& MMT_t^{(\tr)}+ MMT_t^{(\p)} \nonumber \\
  	&=& MMT_t^{<2^j}+ MMT_t^{>2^j}, 
\label{eq:decomposition}
\end{eqnarray}
where $MMT_t^{(\tr)}$ captures the transitory component of the moment computed as a sum of the corresponding elements up to $j$, and $MMT_t^{(\p)}$ captures the persistent component of the moment consisting of the elements larger than $j$. In the next subsections, we show how such decomposition can help us with modelling the persistence of higher moment risks in relation to the time-series of asset returns. Hence from now on, we will refer to transitory and persistent components of higher moment risk within this definition. The two distinct sources of risk will have heterogeneous impact, and our aim is to find how these risks are priced in the cross-section of stocks.

\subsection{Realized moments}
\label{subsec:mkt}

The discussion assumes that higher moments evolve dynamically, but at the same time we need to realize that higher moments are generally hard to measure. In this subsection, we provide brief summary of high-frequency based estimation of higher moment risk measures that we plug into the model. We rely on recent advances in high-frequency econometrics to measure the realized volatility, realized skewness and realized kurtosis and then decompose their fluctuations to transitory and persistent parts so we define heterogeneously persistent higher moments. 

We compute the daily higher moments $MMT_t$ by the means of the realized measures \citep{andersen2001distribution}. The daily realized variance (RDV), realized skewness (RDS), and realized kurtosis (RDK) representing the second, third, and fourth moment of daily returns distribution can be computed from 5-minute prices. Using already well-known arguments of \cite{andersen2001distribution,andersen2003modeling} realized variance can be constructed as sum of the squared high-frequency intraday returns as
\begin{equation}
\label{eq:realized_variance}
RDV_t = \sum_{j = 1}^{K} r_{t, k}^2,
\end{equation}
where $r_{t, k} = p_{t, k / K} - p_{t, (k - 1) / K}$ with $p_{t, k/K}$ denoting a natural logarithm of k-th intraday price on day $t$. We use five-minute returns so that  in $6.5$ trading hours we have $K=78$ intraday returns. Realized Volatility (RDVOL) is computed as $RDVOL_t = \sqrt{RDV_t}$. 

Since we are mainly interested in measuring asymmetry and higher order moments of the daily returns' distribution, we construct a measure of ex-post realized skewness based on intraday returns standardized by the realized variance following \cite{amaya2015} as 
\begin{equation}
\label{eq:realized_skewness}
RDS_t = \frac{\sqrt{N} \sum_{k = 1}^{K} r_{t, K}^3}{RDV_t^{3/2}}.
\end{equation}
The negative values of the realized skewness indicate that stock's return distribution has a left tail that is fatter than the right tail, and positive values indicate the opposite. In addition, extremes of the return distribution can be captured by realized kurtosis
\begin{equation}
\label{eq:realized_kurtosis}
RDK_t  = \frac{N \sum_{k = 1}^{K} r_{t, K}^4} {RV_t^2}.
\end{equation}
Note that as discussed by \cite{amaya2015}, with increasing sampling frequency $K$ realized skewness in the limit separates the jump contribution from the continuous contribution to cubic variation and it captures mainly jump part. This feature is important to note since the measure does not capture leverage effect arising from correlation between return and variance innovations. Hence assets with positive jumps on average will have a positive realized third moment and vice versa, and higher moments measured by high frequency data are likely to contain different information from those computed from daily data or options (see \cite{amaya2015} for rigorous discussion).

\subsection{Average idiosyncratic higher moments}

We define the average idiosyncratic higher moments as the cross-sectional averages of the corresponding realized moments computed for the individual stocks. Since our sample consists of an unbalanced panel, we use rolling windows to obtain the measures of average idiosyncratic moments. Specifically, for each year $y$ in our sample period, we take all the stocks with a complete history of daily returns in the corresponding year, and compute $RDM_{t, i}$, where $RDM \in \{ RDVOL, RDS, RDK \} $ for each stock $i$ and each day $t$ in year $y$. The average idiosyncratic moment for day $t$ is obtained as
\begin{equation}
RDM_t^{(I)} = \frac{1}{N} \sum_{i = 1}^{N} RDM_{t, i},
\end{equation}
where $i$ represent stock $i$ and $N$ is the number of stocks with complete history of daily returns in year $y$. We repeat the procedure for years $y \in \{ 1, ..., Y \}$ where $Y$ is the total number of years in our sample period. The heterogeneously persistent average idiosyncratic higher moments $RDM_t^{(I, h)}$ are obtained analogously as
\begin{equation}
RDM_t^{(I, h)} = \frac{1}{N} \sum_{i = 1}^{N} RDM_{t, i}^{(h)},
\end{equation}
where $h \in \{ \tr, \p \}$.

\subsection{Modelling persistence from high-frequency returns}

The realized higher-order moments exhibit strong time series dynamics \citep{amaya2015} and may thus have unexplored transitory and persistent components that create heterogeneous types of risks matching our theoretical expectation. To explore such risks and work with the assumption of heterogeneous investment horizons, we decompose the realized measures to their horizon-specific components. 

Above we discuss how to decompose an higher moment $MMT_t$ to two orthogonal components capturing economic cycles shorter than $2^j$ periods (for example months) and longer than $2^j$ periods for $j \ge 1$ capturing transitory and permanent information contained in the higher moments respectively. A realized higher moment $RDM_t \in \{RDVOL_t, RDS_t,RDK_t\}$ can be decomposed in an identical fashion as
\begin{eqnarray}
 RDM_t 	&=& RDM_t^{(\tr)}+ RDM_t^{(\p)},  \nonumber \\ 
  		&=& RDM_t^{<2^j}+ RDM_t^{>2^j}, 
\label{eq:decomposition}
\end{eqnarray}
where $RDM_t^{(\tr)}$ denotes transitory component of realized moment computed as a sum of the corresponding elements up to $j$, and $RDM_t^{(\p)}$ denotes persistent component of realized moment consisting of the elements larger than $j$.

A natural question is how can we infer persistence structure of higher moments from realized measures based on high-frequency data? Here we motivate the approach with a simple data generating process to support that the decomposition of realized measures captures the persistence structure of data.  Let's assume returns $r_{t,n}$ evolving over $t \in \{1, \dots, T \}$ days with two components having different level of persistence as
\begin{equation}
r_{t, n} = \underbrace{\beta^{(\tr)} \epsilon_{t, n}^{(\tr)}}_{r_{t, n}^{(\tr)}} + \underbrace{\beta^{(\p)} \epsilon_{t, n}^{(\p)}}_{r_{t, n}^{(\p)}},
\label{eq:ret_simulation}
\end{equation}
with $n \in \{1, \dots, N \}$ denoting the n-th observation on day $t$, $r_{t, n}^{(\tr)}$ and $r_{t, n}^{(\p)}$ denoting high-frequency transitory and persistent components of returns respectively, and $\epsilon_t^{(\tr)} \sim SGT \left( 0, \sigma_{t-1}^{(\tr)}, \lambda_{t-1}^{(\tr)}, \mathcal{P} \right)$, and $\epsilon_t^{(\p)} \sim SGT \left( 0, \sigma_{t-1}^{(\p)}, \lambda_{t-1}^{(\p)}, \mathcal{P}  \right)$ having Skewed generalized t-distribution with  $\sigma_{t-1}^{(.)}$ and $\lambda_{t-1}^{(.)}$ being variance and skewness parameters respectively and $\mathcal{P}$ captures the parameters $p$ and $q$.\footnote{The individual variance and skewness processes are formalized in Appendix~\ref{sec:simulation}. The $p$ and $q$ parameters are fixed with $p = 2$ and $q \to \infty$.}
\begin{figure}[t]
	\begin{center}
		\includegraphics[width=\textwidth]{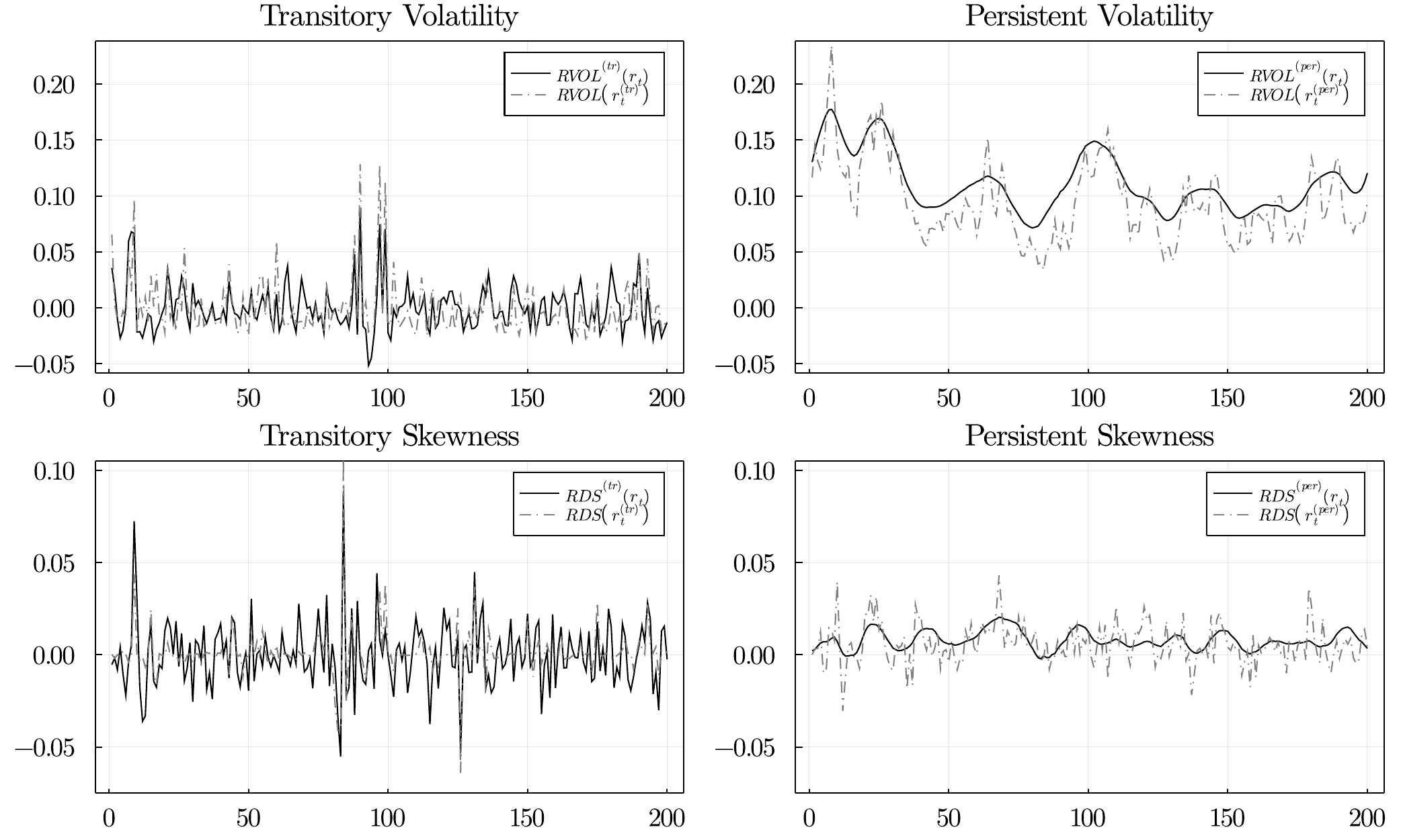}
		\caption[]{Decomposition of Realized Moments vs Realized Moments of Decomposed Returns}
\begin{minipage}{\textwidth} 
\footnotesize
The Figure compares persistence components of realized volatility (RDVOL) and skewness (RDS) estimated on aggregate returns $r_{t, n}$ with volatility and skewness of transitory and persistent returns $r_{t, n}^{(\tr)}$ and $r_{t, n}^{(\p)}$ where $r_{t, n} = r_{t, n}^{(\tr)} + r_{t, n}^{(\p)}$. Transitory and persistent components of volatility and skewness computed on aggregate returns are depicted by solid line (RDVOL$^{(\tr)}(r_t)$, RDVOL$^{(\p)}(r_t)$, RDS$^{(\tr)}(r_t)$, RDS$^{(\p)}(r_t)$ and volatility and skewness of decomposed returns are depicted by dashed line (RDVOL$(r^{(\tr)})$, RDVOL$(r^{(\p)})$, RDS$(r^{(\tr)})$, RDS$(r^{(\p)})$)
    \end{minipage}
\vspace{\medskipamount}
		\label{fig:volatility_simulation}
	\end{center}
\end{figure}
We simulate a realization of this process with $N=288$\footnote{This corresponds to a 5-minute returns data in an 24-hour trading day.} intraday returns over $T=200$ days. First, we compute the realized volatility and skewness on the aggregate returns $r_{t, n}$ and decompose these to transitory and persistent parts as we will do later on real data. Second, we compute the realized volatility and skewness from transitory and persistent components of  returns $r_{t, n}^{(\tr)}$ and $r_{t, n}^{(\p)}$ that are of interest, but are usually hidden from an observer. If our decomposition identifies the persistence components of the observable moments on aggregate returns well, it should match the moments of the persistence return components.

Figure~\ref{fig:volatility_simulation} compares transitory (left) and persistent (right) parts of volatility (top) and skewness (bottom) decomposition. Specifically, we compare components of moments estimated from aggregate returns in solid line to those estimated on true persistence components of returns in dashed line. We can see that transitory and persistent components of realized volatility very closely captures the fluctuations of the volatility of transitory and persistent components of returns. Note that naturally, the persistent component of realized volatility is smoother than the Realized Volatility of persistent returns. Capturing the dynamics of skewness is more difficult since it has substantially lower magnitude but the patterns are captured precisely.

While this is just single realization of the process and estimates have considerable estimation error, this example motivates that transitory and persistent components of realized higher moments capture well the realized moments of transitory and persistent returns and they are a valid approximation for the empirical analysis.

\subsection{Parametrization of the equity premium}
\label{subsec:model}

Above, we motivate why persistence in the higher moments risks is an important aspect of asset pricing and how we can use high-frequency data to compute measures capturing the higher moments of transitory and persistent components of returns. We connect these concepts to the expected asset returns via the Euler equation \citep{hansen1991implications} which determines that assets can be priced using a stochastic discount factor (SDF) denoted by $M_{t+1}$. Based on the discussion, our pricing kernel should incorporate market and idiosyncratic higher moments risks, and should allow for decomposition of these risks to elements operating at different levels of persistence. 

Generally, results of many multi-factor or nonlinear pricing kernel\footnote{Terms stochastic discount factor and pricing kernel both refer to $M_{t+1}$ from the Euler equation (see Equation (\ref{eq:kernel0})), and we treat them as interchangeable.} models stem from arbitrary assumptions about the utility function \citep{dittmar2002}. Making these assumptions is not necessary as preferences over higher moments can be motivated by already existing concepts while maintaining the model-free approach, e.g. motivating the preference for positive skewness via decreasing risk premia in wealth  \citep{arditti1967}. The set of conditions necessary to eliminate counterintuitive risk taking by investors consists of risk aversion, decreasing absolute risk aversion and decreasing absolute prudence \citep{dittmar2002}. Pricing kernel satisfying these conditions is comprised of elements representing moments of returns' distribution up to the fourth order and implying that investors have preferences over both skewness and kurtosis. 

To formalize the discussion, we build on \cite{harvey2000,dittmar2002,maheu2013,chabi2012pricing}, and we assume that a general utility function $U(W_{t+1})$\footnote{The restriction we place on the utility function $U(W_{t+1})$ is that its derivatives, $U^{(n)}(W_{t+1})$ for $n \in \{1, 2, 3, 4 \}$, exist, and are finite.} depending on wealth $W_t$ can be accurately approximated by taking a Taylor expansion up to the fourth order \citep{dittmar2002}. Defining $R_{t+1}^w$ as the simple net return on aggregate wealth, we expand $U(W_{t+1})$ around $W_t (1 + C_t)$, where $C_t$ is an arbitrary return $C_t = E_t(R_{t+1}^w)$. The pricing kernel is defined as $M_{t+1} = U'(W_{t+1})/U'(W_{t})$ and it can be approximated by \citep{maheu2005can} 
\begin{eqnarray}
M_{t+1} & \approx & \sum_{n=0}^{3} \frac{U^{(n+1)}(1+C_{t})}{U'(1) n!}(R_{t+1}^{w}-C_{t})^n \nonumber \\ 
& = & g_{0,t + 1} + g_{1,t + 1} (R_{t+1}^{w}-C_{t}) +g_{2,t + 1} (R_{t+1}^{w}-C_{t})^2 + g_{3,t + 1} (R_{t+1}^{w}-C_{t})^3. 
\label{eq:kernel0}
\end{eqnarray}

Further, we assume that the return of asset $i$ can be decomposed to the systemic component and purely idiosyncratic component \citep{jondeau2019average}, thus we can write
\begin{equation}
R_{t + 1, i}^e = R_{t + 1}^{(m)} + R_{t + 1}^{(i)},
\end{equation}
where $R_{t + 1}^{(m)}$ is the systemic (market) component, and $R_{t + 1}^{(i)}$ is the idiosyncratic component
of the return of asset $i$. If we aggregate for all $i$, we can write the return on aggregate wealth as 
\begin{equation}
R_{t + 1} = R_{t + 1}^{(m)} + \frac{1}{N}\sum_{i = 1}^N R_{t + 1}^{(i)} = R_{t + 1}^{(m)} + R_{t + 1}^{(I)},
\label{eq:market_idio1}
\end{equation}
where $R_{t + 1}^{(I)}$ is the idiosyncratic component of the return on aggregate wealth, and $R_{t + 1}^{(m)}$ is the systemic component of the return on aggregate wealth.

In addition, we assume that both excess return of $i$-th asset $R_{t+1, i}^e$ as well as excess return on aggregate wealth $R_{t+1}$ can be decomposed to elements consisting of transitory and persistent fluctuations as
\begin{equation}
R_{t+1} \equiv \sum_{j=1}^{N} R_{t+1}^{(j)} + R_{t+1}^{(\infty)} = R_{t+1}^{(\tr)} + R_{t+1}^{(\p)},
\label{eq:ret_dec0}
\end{equation}
where $R_{t+1}^{(\tr)} = \sum_{j=1}^{J} R_{t+1}^{(j)}$, and $R_{t+1}^{(\p)} = R_{t+1}^{(\infty)} = R_{t+1}^{(j > J)}$ , and choice of J depends on the economic meaning of transitory and persistent fluctuations.\footnote{Note that due to the equivalence in Equation~(\ref{eq:ret_dec0}), the decomposition is not restricted to two horizons. In fact, we are able to construct components from arbitrary number of horizons by splitting the sum in intermediate points.} 

We employ the pricing kernel from Equation~(\ref{eq:kernel0}), and apply the decompositions from Equation~(\ref{eq:market_idio1}) and  Equation~(\ref{eq:ret_dec0}) which yields the following relationship between returns and higher moments\footnote{The complete and detailed derivation of the model is provided in Appendix~\ref{app:technical}.}
\begin{eqnarray}
r_{t, i} & =  & \sum_{r \in \{ m, I \}} \sum_{h\in\{\tr,\p\}} \beta_{t, i, V}^{(r, h)} RDVOL_{t}^{(r, h)} +  \sum_{r \in \{ m, I \}} \sum_{h\in\{\tr,\p\}} \beta_{t, i, S}^{(r, h)} RDS_{t}^{(r, h)} \nonumber \\
& + &  \sum_{r \in \{ m, I \}} \sum_{h\in\{\tr,\p\}} \beta_{t, i, K}^{(r, h)} RDK_{t}^{(r, h)}, 
\label{eq:estimation1}
\end{eqnarray}
and decomposes returns to a linear combination of Relized Volatility ($RDVOL_t^{(r, h)}$), Realized Skewness ($RS_t^{(r, h)}$), and Realized Kurtosis ($RK_t^{(r, h)}$) of returns on aggregate wealth at a transitory and persistent level $h \in \{\tr, \p\}$. Moreover, $r=m$ denotes market (systemic) component of returns on aggregate wealth, while $r=I$ corresponds to the average idiosyncratic component. 

We follow the Fama-Macbeth cross-sectional regression approach \citep{fama1973regression} to uncover heterogeneity in the persistence of the higher moment risks priced in the stocks returns. Equation~(\ref{eq:estimation1}) is used to estimate the first-stage regression coefficients expressing the sensitivities to individual sources of risk. After obtaining the coefficients from Equation~(\ref{eq:estimation1}), we run the following second-stage regression
\begin{eqnarray}
r_{t + 1, i} &  = &   \sum_{r \in \{ m, I \}} \sum_{h\in\{\tr,\p\}}  \lambda_{t, V}^{(r, h)} \beta_{t, i, V}^{(r, h)} +  \sum_{r \in \{ m, I \}} \sum_{h\in\{\tr,\p\}}  \lambda_{t, S}^{(r, h)} \beta_{t, i, S}^{(r, h)}  \nonumber \\ 
& + &  \sum_{r \in \{ m, I \}} \sum_{h\in\{\tr,\p\}}  \lambda_{t, K}^{(r, h)} \beta_{t, i, K}^{(r, h)},
\label{eq:estimation2}
\end{eqnarray}
where $V$ denotes volatility risk, $S$ denotes skewness risk, and $K$ denotes kurtosis risk. We use the rolling windows approach to estimate the coefficients from Equation~(\ref{eq:estimation1}) and Equation~(\ref{eq:estimation2}).
\subsection{Data}

We analyze high-frequency intraday data about all stocks from New York Stock Exchange (NYSE), the American Stock Exchange (AMEX), and NASDAQ included in the CRSP database. We record prices every five minutes starting 9:30 EST and construct five-minute log-returns for the period 9:30 EST to 16:00 EST for a total of 78 intraday returns. We construct the five- minute grid by using the last recorded price within the preceding five-minute period, and we consider excess returns as required by our empirical model outlined later in the text. A ``coarse'' five-minute sampling scheme aims to balance the bias induced by market microstructure effects and mirrors common practice in the literature.

\begin{figure}[t!]
	\begin{center}
		\includegraphics[width=\textwidth]{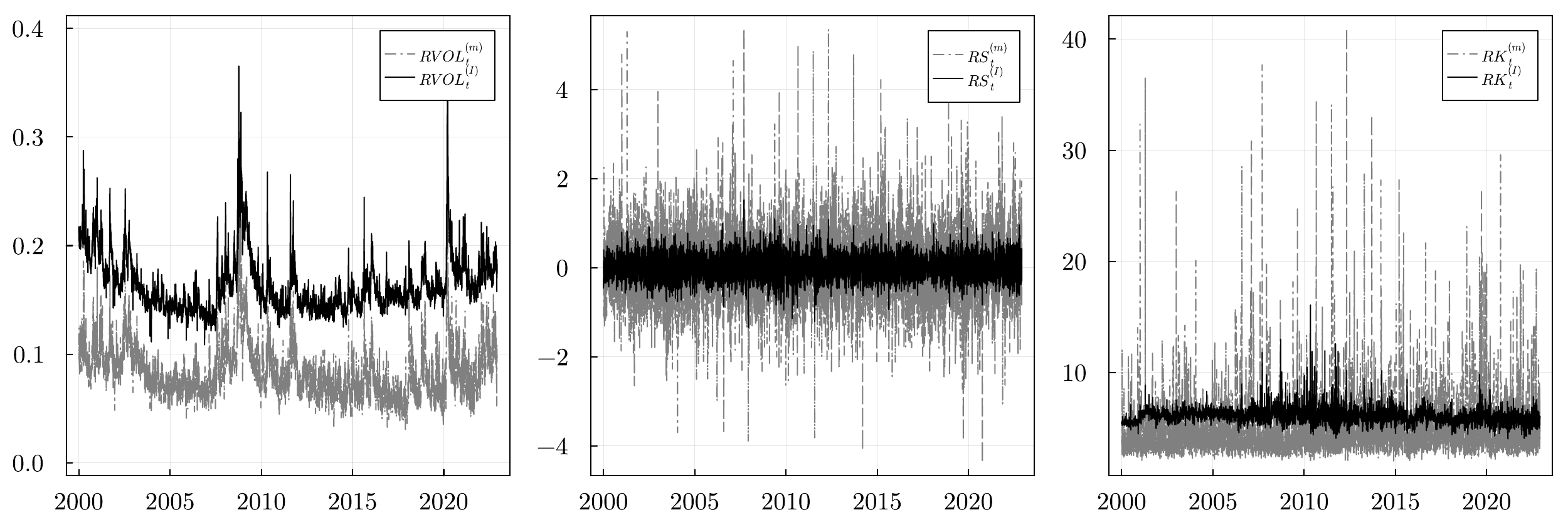}
		\includegraphics[width=\textwidth]{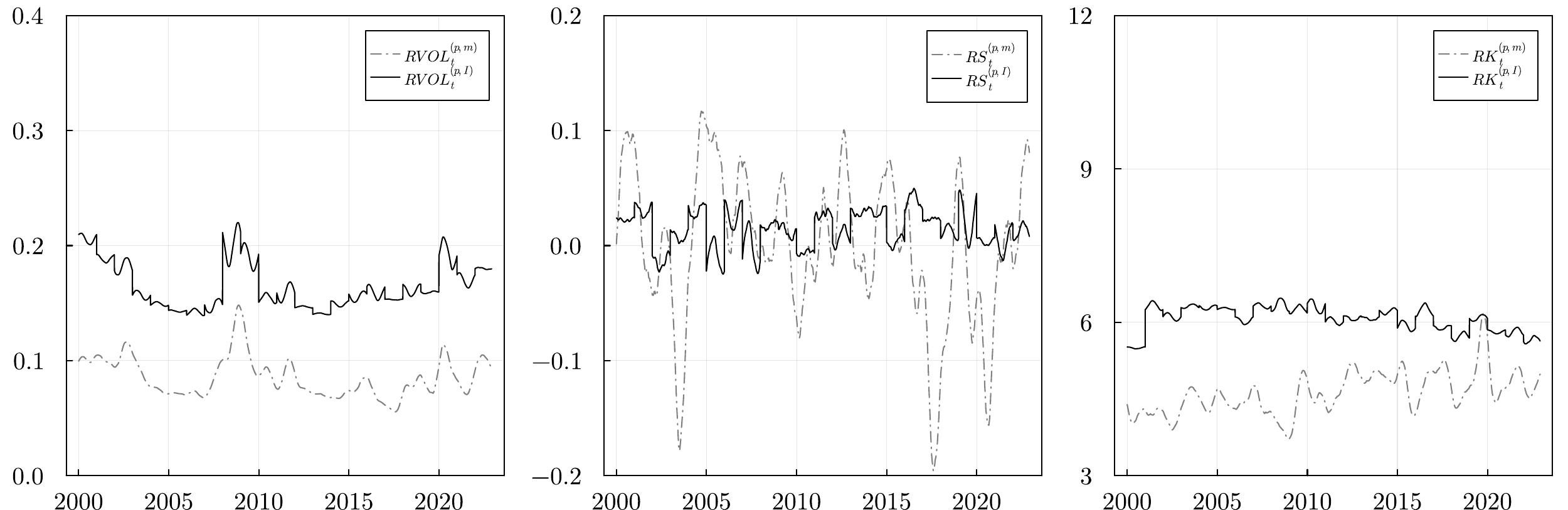}
		\caption{\textbf{Daily realized measures for cross-section of stocks}} 
		\vspace{\medskipamount}
		\begin{minipage}{\textwidth} 
			\footnotesize
			The top row depicts market (dashed) and average idiosyncratic (black line) realized measures of the cross-section of daily stocks returns. The bottom row depicts the long-term components that capture fluctuations of all measures longer than half a year. 
		\end{minipage}
		\label{fig:moments_stocks}
	\end{center}
\end{figure}

After data cleaning, we are left with the cross-section of 12 231 stocks covering the period from 01/2000 to 12/2022.\footnote{The data cleaning contains following steps. For each month $m$ we omit stocks with price less than 1USD (we use 5 USD threshold as a robustness check) on the last trading day of month $m-1$. Then, we also omit stocks whose return fell into the top/bottom percentile of all stocks returns on the last trading day of month $m - 1$.} The daily return of an asset is constructed from dividends and splits adjusted prices as an excess return of a logarithmic difference between opening and closing price for a given period. We use the 3-month Treasury Bill rate as the risk-free rate. We perform the empirical exercise using daily returns as well as daily measures of higher moments risks since the daily frequency is crucial to distinguish the transitory and persistent components and capture especially the transitory fluctuations in the higher moments risks.

Figure \ref{fig:moments_stocks} shows all the computed heterogeneously persistent realized moments for our sample of stocks. The first row contrasts market and average idiosyncratic realized volatility, realized skewness and realized kurtosis of stocks returns. It is visible that the measures of average idiosyncratic realized volatility and realized kurtosis are larger in magnitude than the market ones, and that market realized skewness shows larger fluctuations than average idiosyncratic realized skewness. The bottom row comparing long-term fluctuations of the market and average idiosyncratic measures shows similar dynamics of these two sources of volatility and kurtosis risk with the average idiosyncratic measures being larger in magnitude again. Meanwhile, we observe interestingly different dynamics between the persistent fluctuations of market and average idiosyncratic skewness measures. While market skewness fluctuates around zero with three large negative spikes during 2004, 2017, and 2021, the average idiosyncratic skewness is positive for almost the whole period and the fluctuations it displays are of substantially lower magnitude.

\begin{table}[t!]
\tiny
\centering
\caption[]{Correlations}
\begin{minipage}{\textwidth} 
\footnotesize
This table provides the correlation matrix for the realized market and average idiosyncratic realized moments in the Panel A, correlation matrix for the transitory and persistent realized market and average idiosyncratic moments in the Panel B, and Panel C.
\end{minipage}
\vspace{\medskipamount}

  \begin{tabular}{lrrrrrrrrrrrr}
    \hline\hline 
    \textbf{Panel A } & $RVOL_{t}^{(m)}$ & $RS_{t}^{(m)}$ & $RK_{t}^{(m)}$ & $RVOL_{t}^{(I)}$ & $RS_{t}^{(I)}$ & $RK_{t}^{(I)}$ \\
    \cline{1-7} 
    $RVOL_{t}^{(m)}$ &   &   &   &   &   &   \\
    $RS_{t}^{(m)}$ & -0.01	&   &   &   &   &   \\
    $RK_{t}^{(m)}$ & -0.03	 & 0.117 &   &   &   &   \\
    $RVOL_{t}^{(I)}$ & 0.872 & 0.014 & -0.076	 &   &   &   \\
    $RS_{t}^{(I)}$ & -0.035	& 0.65 & 0.095 & -0.001	 &   & \\
    $RK_{t}^{(I)}$ & -0.105	& 0.115 & 0.298 & -0.104 & 0.122 &  \\
    \\
    \cline{1-7}  
    \textbf{Panel B } & $RVOL_{t}^{(\tr, m)}$ & $RVOL_{t}^{(\p, m)}$ & $RS_{t}^{(\tr, m)}$ & $RS_{t}^{(\p, m)}$ & $RK_{t}^{(\tr, m)}$ & $RK_{t}^{(\p, m)}$ & &   \\
    \cline{1-7}   
    $RVOL_{t}^{(\tr, m)}$ &  & & & & & & & & & & &  \\
    $RVOL_{t}^{(\p, m)}$ & 0.128 &  & & & & & & & & & & \\
    $RS_{t}^{(\tr, m)}$ & -0.025 & 0.003 &  & & & & & & & & &  \\
    $RS_{t}^{(\p, m)}$ & 0.041 & 0.082 & 0.026	&  & & & & & & & &  \\
    $RK_{t}^{(\tr, m)}$ & 0.064 & -0.016 & 0.123 & -0.004&  & & & & & & &  \\
    $RK_{t}^{(\p, m)}$ & -0.075 & -0.6 & -0.002 & -0.236 & 0.042 & &  & & & & & \\
     \cline{1-7}   \cline{9-13}
    \textbf{Panel C} & $RVOL_{t}^{(\tr, m)}$ & $RVOL_{t}^{(\p, m)}$ & $RS_{t}^{(\tr, m)}$ & $RS_{t}^{(\p, m)}$ & $RK_{t}^{(\tr, m)}$ & $RK_{t}^{(\p, m)}$ & & $RVOL_{t}^{(\tr, I)}$ & $RVOL_{t}^{(\p, I)}$ & $RS_{t}^{(\tr, I)}$ & $RS_{t}^{(\p, I)}$ & $RK_{t}^{(\tr, I)}$  \\
    \cline{1-7}   \cline{9-13}
    $RVOL_{t}^{(\tr, I)}$ & 0.804 & 0.252 & 0.01 & 0.086 & -0.001 & -0.134 & & & & & & \\
    $RVOL_{t}^{(\p, I)}$ & 0.126 & 0.82 & 0.005 & 0.01 & -0.018 & -0.441 & & 0.126 & & & & \\
    $RS_{t}^{(\tr, I)}$ & -0.037 & -0.004 & 0.653 & 0.015 & 0.096 & 0 &  & 0 & 0.001 & & & \\
    $RS_{t}^{(\p, I)}$ & -0.016 & -0.149 & 0.014 & 0.104 & 0.008 & 0.096 & & 0.006 & -0.03 & 0.03 & & \\	
    $RK_{t}^{(\tr, I)}$ & -0.11 & -0.026 & 0.129 & -0.01 & 0.324 & 0.039 & & -0.011 & -0.016 & 0.134 & 0.005 & \\
    $RK_{t}^{(\p, I)}$ & -0.074 & 0.007 & -0.003 & -0.002 & 0.02 & -0.095 & & -0.047 & -0.293 & 0 & 0.008 & 0.039 \\  
    \hline\hline
  \end{tabular}

\label{tab:corr}
\end{table}

Table~\ref{tab:corr} further reveals that market realized volatility is strongly correlated with average idiosyncratic volatility while the relation is weaker for the realized skewness and almost disappears in case of kurtosis. This suggests that the average idiosyncratic and market parts of the third and fourth moments carry substantially different information. More importantly, Panel B shows that transitory and persistent components of market higher moments are generally uncorrelated except strong negative correlation between persistent components of market volatility and kurtosis. 

The pattern is not very different in the right part of Panel C where correlation matrix of heterogeneously persistent components of average idiosyncratic moments is displayed. While most of the terms are not correlated, the persistent components of average idiosyncratic volatility and kurtosis show mild negative correlation. Finally, left part of Panel C displaying the cross-correlations between the heterogeneously persistent components of market and average idiosyncratic moments confirms the previous findings. The values on the diagonal document that correlation between the corresponding components of the same moment  is strongest in case of volatility, and becomes weaker in case of skewness and especially kurtosis. Moreover, we observe negative correlation between persistent components of average idiosyncratic volatility and market kurtosis.

\section{Heterogeneously Persistent Higher Moments and Expected Returns}
\label{sec:results}

The preliminary analysis from the previous section documents that various types of risks and information are hidden in the transitory and persistent components of higher moments. Moreover, it implies that individual assets exhibit different exposures to the heterogeneously persistent components of the higher moments risks, and that there are two relevant sources of such risk; market and idiosyncratic. Subsequently, different exposures to the corresponding risk factors yield different returns on average. 

We exploit these findings using Fama-Macbeth type cross-sectional regressions \citep{fama1973regression} to evaluate the ability of persistent and transitory higher moments risks given by Equation~(\ref{eq:estimation1}) and Equation~(\ref{eq:estimation2}) to predict subsequent returns. The first-stage regression has the following general form
\begin{equation}
r_{t, i} = \sum_{k = 1}^K \beta_{t, i, k} X_{t, k} + \sum_{j = 1}^J \gamma_{t, i, j} Z_{t, j}
\label{eq:model_stage1}
\end{equation}
where $X_{t, k}$ is the k-th higher moment variable at time t, and $Z_{t, j}$ is the j-th control variable at time t. The second-stage regression is then
\begin{equation}
r_{t + 1, i} = \sum_{k = 1}^K \lambda_{t, k} \beta_{t, i, k} + \sum_{j = 1}^J \phi_{t, j} \gamma_{t, i, j}.
\label{eq:model_stage2}
\end{equation}
The individual models differ in the content of $\mathbf{X} = \{ X_1, \dots, X_K \}$, and $\mathbf{Z} = \{ Z_1, \dots, Z_J \}$. We employ rolling windows approach to estimate Equation~(\ref{eq:model_stage1}) and Equation~(\ref{eq:model_stage2}). Specifically, we use 6 months of daily returns\footnote{Complete history of returns in the past 6 months is required for stock $i$ to be included.} to estimate the $\beta_{t, i, k}$ and $\gamma_{t, i, j}$ coefficients from Equation~(\ref{eq:model_stage1}) for each $i$, $k$, and $j$. Then, we use these coefficients to estimate Equation~(\ref{eq:model_stage2}), where we use average daily returns over the next week\footnote{We use Wednesday through Tuesday as a definition of the week. One valid observation in the week $t + 1$ is sufficient for stock $i$ to be included.} as an explained variable. Once the $\lambda_{t, k}$ and $\phi_{t, j}$ coefficients are estimated, we roll one week forward and repeat the procedure.

The aim of this paper is to show that decomposing higher moments risks to the transitory and persistent components is a crucial step towards understanding how these risks are priced in the cross-section of stocks returns. Hence, we present models considering the aggregate measures of higher moments risks and models considering heterogeneously persistent components of the higher moments risks separately. The full list of model specifications corresponding to Table~\ref{tab:crsp6m1usd} is provided in Appendix~\ref{app:models}.
    
The estimated models combine the effects of volatility, skewness, and kurtosis representing different aspects of risk on the financial markets. Generally, investors with sensible preferences over risk prefer portfolios with lower volatility, higher skewness, and lower kurtosis \citep{kimball1993standard}, hence need to be compensated by higher returns for accepting portfolios with higher volatility, lower skewness, or higher kurtosis. However, it is not straightforward to connect these phenomena to the prices of risk. We are able to capture the conditions under which the price of volatility is positive/negative thanks to the Intertemporal CAPM \citep{merton1973intertemporal, campbell1996understanding, ang2006, chen2002intertemporal}. 

Market volatility is priced because it serves as a hedge against future changes on the market. If high volatility is connected to downward price movements, then an asset whose return has a positive sensitivity to market volatility is a desirable hedging instrument, hence the negative price of market volatility in such case. If the opposite holds, an asset with positive sensitivity to market volatility is undesirable, hence investors should require compensation for holding such asset. The sign of market volatility risk price is therefore usually expected to be negative due to the presence of the leverage effect. Empirically, there is evidence that innovations to market volatility are negatively priced in the cross-section of asset returns \citep{ang2006, chang2013}, and non-robust evidence that idiosyncratic volatility has a positive effect on subsequent returns on assets in the TAQ database \citep{bollerslev2018}.

Price of higher moments risk cannot be determined by observing the empirical correlations, since such approach would ignore the individual investors' risk attitudes like skewness preference. \cite{chabi2012pricing} concludes that the prices of market skewness risk, and market kurtosis risk depend on the fourth and fifth derivative of the utility function which are essentially hard to sign. Hence, we shall perceive determining the prices of higher moments risks merely as an empirical exercise. Lastly, the transitory and persistent components should be generally priced with the same sign as the corresponding aggregate risks. However, it can not be ruled out that the decomposition will uncover some effects that are opposite to those prevailing on aggregate for the particular higher moment.

The empirical results provide several new insights. We are able to disentangle how volatility, skewness and kurtosis are priced in the cross-section of stocks. Considering the transitory and persistent components of these risks separately reveals the types of fluctuations that are mostly relevant to the investors with regard to the particular higher moments risks. Decomposing the risks into transitory and persistent components should be a key feature of an asset pricing model \citep{neuhierl2021frequency} with potential to help explain the puzzles posed by previous empirical tests. Controlling for market as well as average idiosyncratic risk contributes to the recent debate of how these two sources risk are priced individually as well as jointly.

\subsection{Empirical evidence}

Below we report the results of the cross-sectional regressions described above using daily stock data and predicting average daily returns over the next week. The main results using a 6-month rolling window and a USD 1 threshold in the stock data filtering are presented in Table~\ref{tab:crsp6m1usd}. The models incorporating aggregate measures of higher moment risk are presented in Panel A, and the models incorporating heterogeneously persistent measures of higher moment risk are presented in Panel B. To uncover the relationship between the market and average idiosyncratic components of risk, we present results from models incorporating only market risk\footnote{Note that market returns are approximated by S\&P 500 returns.} (model \textbf{1} and model \textbf{4}) and only idiosyncratic risk (model \textbf{2} and model \textbf{5}). The columns labelled \textbf{3} and \textbf{6} present the full models including Fama-French 3-factor (FF3) control variables.\footnote{The role of the FF3 control variables (MKT, SMB, HML) is to observe whether the significance of the coefficients in our models is robust to controlling for the financial market anomalies identified. For the sake of clarity, we do not present results for these coefficients.}

\begin{table}[p!]
\footnotesize
\centering
\caption[]{Heterogeneously persistent higher moments risks and the cross-section of stocks}
\begin{minipage}{\textwidth} 
\footnotesize
We report the lambda coefficients from the second-stage regression specified by Equation~(\ref{eq:model_stage2}). The prices of risk specified in Equation~(\ref{eq:model_stage1}) are estimated using daily stocks data. The full list of model specifications is provided in Appendix~\ref{app:models}. We use 1 USD threshold in the data cleaning procedure. The t-statistics displayed in parentheses are computed using Newey-West standard errors.
\end{minipage}
\vspace{\medskipamount}

\begin{tabular}{l c c c c c c }
\hline
\hline
\textbf{Variable} & \textbf{1} & \textbf{2} & \textbf{3} & \textbf{4} & \textbf{5} & \textbf{6} \\
\hline
const & 0.000281 & 0.000322 & 0.000358 & 0.000297 & 0.00032 & 0.000356 \\  & (2.181111) & (2.598543) & (3.182271) & (2.341864) & (2.662696) & (3.168651) \\
$RVOL_{t}^{(m)}$ & -0.00012 &  & -0.000128 &  &  &  \\  & (-1.757531) &  & (-2.121501) &  &  &  \\
$RS_{t}^{(m)}$ & -0.009586 &  & -0.001832 &  &  &  \\  & (-1.04716) &  & (-0.289437) &  &  &  \\
$RK_{t}^{(m)}$ & -0.004844 &  & 0.00809 &  &  &  \\  & (-0.23016) &  & (0.458247) &  &  &  \\
$RVOL_{t}^{(m, \tr)}$ &  &  &  & -0.000139 &  & -0.00012 \\  &  &  &  & (-2.078907) &  & (-2.160449) \\
$RVOL_{t}^{(m, \p)}$ &  &  &  & -0.000004 &  & -0.000012 \\  &  &  &  & (-0.205649) &  & (-0.686443) \\
$RS_{t}^{(m, \tr)}$ &  &  &  & -0.008255 &  & -0.001681 \\  &  &  &  & (-0.949761) &  & (-0.279828) \\
$RS_{t}^{(m, \p)}$ &  &  &  & 0.000149 &  & 0.000029 \\  &  &  &  & (0.475149) &  & (0.102458) \\
$RK_{t}^{(m, \tr)}$ &  &  &  & -0.015453 &  & 0.005208 \\  &  &  &  & (-0.806734) &  & (0.319513) \\
$RK_{t}^{(m, \p)}$ &  &  &  & -0.000689 &  & -0.000665 \\  &  &  &  & (-0.291965) &  & (-0.32497) \\
$RVOL_{t}^{(I)}$ &  & -0.000181 & -0.000201 &  &  &  \\  &  & (-1.956457) & (-2.142203) &  &  &  \\
$RS_{t}^{(I)}$ &  & -0.003024 & -0.000952 &  &  &  \\  &  & (-1.005168) & (-0.432845) &  &  &  \\
$RK_{t}^{(I)}$ &  & -0.004833 & -0.00115 &  &  &  \\  &  & (-0.917668) & (-0.251865) &  &  &  \\
$RVOL_{t}^{(I, \tr)}$ &  &  &  &  & -0.00019 & -0.000179 \\  &  &  &  &  & (-2.146234) & (-2.047274) \\
$RVOL_{t}^{(I, \p)}$ &  &  &  &  & -0.000016 & -0.000032 \\  &  &  &  &  & (-0.414952) & (-0.954213) \\
$RS_{t}^{(I, \tr)}$ &  &  &  &  & -0.002961 & -0.000759 \\  &  &  &  &  & (-1.022626) & (-0.356827) \\
$RS_{t}^{(I, \p)}$ &  &  &  &  & -0.000196 & -0.000187 \\  &  &  &  &  & (-2.044242) & (-2.082629) \\
$RK_{t}^{(I, \tr)}$ &  &  &  &  & -0.00355 & -0.000535 \\  &  &  &  &  & (-0.736045) & (-0.129823) \\
$RK_{t}^{(I, \p)}$ &  &  &  &  & -0.00099 & -0.000408 \\  &  &  &  &  & (-0.805759) & (-0.358429) \\
\hline
\end{tabular}

\label{tab:crsp6m1usd}
\end{table}

Panel A, estimated under the assumption of homogeneous preferences of investors over different investment horizons, shows how the aggregate shocks to the individual higher moments are priced into the cross-section of stocks. We find that average idiosyncratic volatility is priced into the cross-section of stock returns, while the evidence on market volatility is not straightforward. Specifically, the coefficient corresponding to idiosyncratic volatility is $-0.000181$ with a significant t-statistic of $-1.96$ and $-0.000201$ with a significant t-statistic of $-2.14$ in models \textbf{2} and \textbf{3} of Table~\ref{tab:crsp6m1usd} respectively. Table~\ref{tab:crsp6m5usd} supports the evidence regarding average idiosyncratic volatility, as it is associated with significant t-statistics of $-1.75$ and $-2.03$ respectively. Meanwhile, the coefficients associated with market volatility have significant t-statistics in Table~\ref{tab:crsp6m1usd}, but we do not find any evidence to support the robustness of this finding in Table~\ref{tab:crsp6m5usd}.

The negative sign of the coefficients associated with average idiosyncratic volatility suggests that investors are willing to accept lower returns for stocks with higher exposure to average idiosyncratic volatility, i.e. they prefer such stocks. Such investor preferences are consistent with previous empirical findings \citep{chang2013, ang2006} as well as the discussion of the signs of the individual higher moments above. The importance of market or idiosyncratic volatility in predicting equity risk premia has been documented in the literature, but average idiosyncratic volatility has received less attention.\footnote{In contrast, the importance of average idiosyncratic skewness has been documented by \cite{jondeau2019average} or \cite{langlois2020measuring}.} We show that average idiosyncratic volatility is priced into the cross-section of stock returns while controlling for market volatility, i.e. average idiosyncratic volatility contains different information than market volatility and has power in predicting stock risk premia. Moreover, average idiosyncratic volatility affects future returns with the same sign as suggested by theory and shown by previous empirical results, which contributes to the notion that investors have a preference for volatility for hedging purposes. Market volatility captures the volatility risk of large cap stocks, while average idiosyncratic volatility also captures the volatility risk of mid cap and small cap stocks. Hedging against the price movements of such stocks becomes more important as financial markets become more asymmetrically interconnected.

In contrast to volatility, neither Table~\ref{tab:crsp6m1usd} nor Table~\ref{tab:crsp6m5usd} shows evidence of a statistically significant relationship between exposure to skewness or kurtosis risk and subsequent stock returns. Such a finding contrasts with previous empirical findings, as exposure to market skewness has been shown to be negatively priced in the cross-section of stocks \citep{kraus1976, harvey2000, chang2013}. Moreover, average skewness has also been shown to be a significant predictor of returns in financial markets \citep{jondeau2019average, langlois2020measuring}. Our setting is different in that we consider the effects of volatility, skewness and kurtosis, taking into account both market and average idiosyncratic risk. We construct risk factors from these measures and examine how they affect subsequent stock returns. We find no evidence that either source of skewness risk is significantly priced into the cross-section of stocks in such an environment.

Similarly, we find no evidence that either market or average idiosyncratic kurtosis risk is priced into the cross-section of CRSP stocks. The contrast with the literature is less pronounced in the case of kurtosis, as it has not received the same attention as skewness and, in particular, volatility. Moreover, the empirical evidence to date on kurtosis is mixed. \cite{amaya2015} found evidence that idiosyncratic kurtosis positively affects subsequent stock returns, but the results were not robust to model changes. \cite{chang2013} find little evidence that market kurtosis risk is priced into the cross-section of stock returns. In particular, we contribute to the discussion by pointing out that a factor representing average idiosyncratic kurtosis risk is not priced into the cross-section of stock returns, thus confirming that kurtosis plays a minor role in predicting equity risk premia.

The main results are presented in Panel B of Table~\ref{tab:crsp6m1usd} (and Table~\ref{tab:crsp6m5usd}), since the main focus of our paper is on how the transitory and persistent fluctuations of each higher moment affect the cross-section of stock returns. The main contribution of such an approach to the literature is twofold. First, treating higher moment risks as heterogeneously persistent contributes to the debate on the role of these risks themselves in predicting financial asset returns. We argue that the ambiguity of the empirical results associated with some of the moments discussed above stems from the fact that we aggregate these risks across horizons. Looking at the heterogeneously persistent shocks to individual moments separately helps to uncover relationships that would otherwise remain hidden. Second, for each of the higher moments, we identify the degree of persistence that dominates in terms of asset pricing. This allows us to uncover the persistence structure of the fluctuations that play the most important role in the mechanisms that generate risk premia in stock markets, and thus we make a major contribution to the literature on long-term risk \citep[e.g.][]{bansal2004, maheu2007, bandi2023}.

While aggregate measures of skewness are not priced in our sample of stocks, Panel B of Table~\ref{tab:crsp6m1usd} shows that the persistent component of average idiosyncratic skewness is associated with significant changes in risk premia. It has a coefficient of $-0.000196$ with a significant t-statistic of $-2.04$ in model~\textbf{5} and a coefficient of $-0.000187$ with a significant t-statistic of $-2.08$ in model~\textbf{6}, controlling for market risk and FF3 factors. The economic intuition about preferences for skewness is that investors seek extreme positive returns from holding positively skewed assets. In our setting, the negative price of skewness risk expresses a preference for assets with higher exposure to average idiosyncratic skewness risk, i.e. assets that earn high returns when average idiosyncratic conditional skewness is high.\footnote{When we refer to high skewness, we assume that it has a positive value, and vice versa when we refer to low skewness. This is because skewness tends to fluctuate around zero. When skewness is negative, investors will want to avoid assets with high exposure to skewness risk, as these will realise the highest losses in such a market condition. Consequently, assets with high exposure to skewness risk would earn higher expected returns if investors' preferences are associated with the extreme negative returns.}

The case of skewness risk shows not only that decomposing shocks into elements of different persistence is a crucial step in deep understanding  the role of risks arising from each higher moment, but also the difficulty of determining the term structures of these risks. Intuition as well as previous results suggest that the term structure of skewness risk should be downward sloping \citep{neuberger2019skewness}. However, our results show that investors tend to prefer the persistent component of skewness risk, indicating an upward-sloping term structure. Thus, our results reveal that investors' preferences for skewness are not entirely driven by the demand for lottery-like assets and that we should not treat skewness risk as an exclusively short-term phenomenon.

We further demonstrate the importance of uncovering the persistence structure of individual risks using market volatility as an example. As reported above, we find evidence that market volatility risk is priced into the cross-section of stock returns, but we cannot confirm the significance of such an effect through Table~\ref{tab:crsp6m5usd}. If we consider market volatility risk as a combination of transitory and persistent components, we find that the transitory components are robustly priced in both Table~\ref{tab:crsp6m1usd} and Table~\ref{tab:crsp6m5usd}. The coefficient associated with the transitory component of market volatility is $-0.000139$ with a significant t-statistic of $-2.08$ and $-0.00012$ with a significant t-statistic of $-2.16$ in models \textbf{4} and \textbf{6} of Table~\ref{tab:crsp6m1usd} respectively. Table~\ref{tab:crsp6m5usd} suggests a qualitatively similar effect of transitory shocks to market volatility on subsequent stock returns.

Average idiosyncratic volatility shows a robust negative effect on subsequent stock returns, even after controlling for aggregate risk. The decomposition into heterogeneously persistent components suggests that average idiosyncratic risk is a purely transitory phenomenon. The coefficients associated with transitory average idiosyncratic risk are $-0.00019$ with a significant t-statistic of $-2.15$ and $-0.000179$ with a significant t-statistic of $-2.04$ in models \textbf{5} and \textbf{6} of Table~\ref{tab:crsp6m1usd}, respectively. In contrast, we find no evidence that the persistent components of average idiosyncratic volatility risk are priced into the cross-section of stock returns.

The transitory nature of volatility risk is a novel contribution to the literature, which predominantly emphasises the importance of long-run risk both overall \citep{bansal2004, neuhierl2021frequency, bandi2023} and in the case of volatility \citep{maheu2007, kim2013}. While \cite{adrian2008} find that both the short-run and the long-run components are priced into the cross-section of stock returns, we extend this notion to average idiosyncratic volatility risk, suggesting that investors in the US stock markets show a preference for the transitory fluctuations of volatility.

Finally, we find no evidence that either transitory or persistent components of market and average idiosyncratic kurtosis are priced in the cross-section of stocks, consistent with Panel A. Our results suggest that the failure to robustly show that kurtosis is priced in the financial cross-sections \citep{chang2013, amaya2015} is not due to the perception that the shocks to kurtosis are homogeneously persistent.

Overall, the results in Panel B show that the degree of persistence is an important property of the higher moment risks and that it is not trivial to assess the term structures of the individual risks. The decomposition into components with different degrees of persistence appears to be crucial for uncovering how average idiosyncratic skewness and market volatility are priced into the cross-section of stock returns. We also document that investors show preferences only for the transitory component of market and average idiosyncratic volatility, while preferences for average idiosyncratic skewness risk are associated only with the persistent shocks.

\section{Conclusion}
\label{sec:concl}

We show that transitory and persistent higher moment risks are important for the cross-section of stock returns. Short- and long-term fluctuations in realized market and average idiosyncratic volatility, as well as skewness and kurtosis, are priced differently in the cross-section of asset returns, implying a heterogeneous persistence structure of different sources of higher moments risks. We also show that market and average idiosyncratic higher moments risks carry different information and that they enter investors' decision making separately.

Our empirical results have implications along several dimensions. First, they highlight the importance of disentangling the persistence structures of each higher moment in order to fully understand how their fluctuations affect subsequent stock returns. Specifically, we find no significant effects when looking at aggregate measures of market volatility and average idiosyncratic skewness. However, treating these risks as heterogeneously persistent allows us to uncover how they are priced in the cross-section of stocks. Transient shocks to market volatility are priced by investors, while they also show preferences for persistent fluctuations in average idiosyncratic skewness.

Second, the decomposition of risks into heterogeneously persistent components allows us to assess the term structures of these risks. Thus, we contribute to the debate on how transitory and persistent risks affect risk premia in financial markets, which mostly emphasises the importance of long-run risks \citep{bansal2004, maheu2007, kim2013, neuhierl2021frequency, bandi2023}. On the one hand, we find that both market and average idiosyncratic volatility are priced as transitory risks in the cross-section of stocks. On the other hand, we find an upward sloping term structure of average idiosyncratic skewness risk. Overall, our results suggest that volatility risk should be perceived as transitory, while skewness risk should be perceived as persistent.

We also contribute to the debate on the importance of idiosyncratic risks over and above market risks. We can no longer assume that idiosyncratic risks can be fully diversified away as global financial markets become increasingly interconnected and local shocks can spread very quickly around the world. We document that the average idiosyncratic volatility risk is priced into the cross-section of stocks more than the market volatility risk, and that it is the more important of these two sources of risk. Decomposing the higher moment risks into the heterogeneously persistent components also provides evidence that average idiosyncratic skewness risk has a greater weight in predicting equity risk premia than market skewness risk. Thus, we indirectly confirm that average skewness is also important \citep{jondeau2019average, langlois2020measuring} and we extend this notion to average volatility. Finally, consistent with the results in the literature, we find no evidence that market or average idiosyncratic kurtosis is priced into the cross-section of stocks. Moreover, we find no evidence that the failure to document significant effects of kurtosis risk on subsequent stock returns is due to the assumption of a homogeneous persistence structure of this risk.

Overall, the results of the paper provide new insights into the sources of asset predictability with respect to the decomposition of higher moment risks. We attempt to rationalise our findings in a formal theoretical model.

\clearpage

\begingroup
\linespread{1}
\setlength{\bibsep}{0pt}
\setlength{\bibhang}{1.0em}
\bibliographystyle{chicago}
\bibliography{BIBLIOGRAPHY}

\begin{thebibliography}{}

\bibitem[\protect\citeauthoryear{Adrian and Rosenberg}{Adrian and
  Rosenberg}{2008}]{adrian2008}
Adrian, T. and J.~Rosenberg (2008).
\newblock Stock returns and volatility: Pricing the short-run and long-run
  components of market risk.
\newblock {\em The Journal of Finance\/}~{\em 63\/}(6), 2997--3030.

\bibitem[\protect\citeauthoryear{Agarwal, Bakshi, and Huij}{Agarwal
  et~al.}{2009}]{agarwal2009}
Agarwal, V., G.~Bakshi, and J.~Huij (2009).
\newblock Do higher-moment equity risks explain hedge fund returns?
\newblock {\em Robert H. Smith School Research Paper No. RHS\/}, 06--153.

\bibitem[\protect\citeauthoryear{Amaya, Christoffersen, Jacobs, and
  Vasquez}{Amaya et~al.}{2015}]{amaya2015}
Amaya, D., P.~Christoffersen, K.~Jacobs, and A.~Vasquez (2015).
\newblock Does realized skewness predict the cross-section of equity returns?
\newblock {\em Journal of Financial Economics\/}~{\em 118\/}(1), 135--167.

\bibitem[\protect\citeauthoryear{Andersen, Bollerslev, Diebold, and
  Labys}{Andersen et~al.}{2001}]{andersen2001distribution}
Andersen, T.~G., T.~Bollerslev, F.~X. Diebold, and P.~Labys (2001).
\newblock The distribution of realized exchange rate volatility.
\newblock {\em Journal of the American Statistical Association\/}~{\em
  96\/}(453).

\bibitem[\protect\citeauthoryear{Andersen, Bollerslev, Diebold, and
  Labys}{Andersen et~al.}{2003}]{andersen2003modeling}
Andersen, T.~G., T.~Bollerslev, F.~X. Diebold, and P.~Labys (2003).
\newblock Modeling and forecasting realized volatility.
\newblock {\em Econometrica\/}~{\em 71\/}(2), 579--625.

\bibitem[\protect\citeauthoryear{Ang, Hodrick, Xing, and Zhang}{Ang
  et~al.}{2006}]{ang2006}
Ang, A., R.~J. Hodrick, Y.~Xing, and X.~Zhang (2006).
\newblock The cross-section of volatility and expected returns.
\newblock {\em The Journal of Finance\/}~{\em 61\/}(1), 259--299.

\bibitem[\protect\citeauthoryear{Arditti}{Arditti}{1967}]{arditti1967}
Arditti, F.~D. (1967).
\newblock Risk and the required return on equity.
\newblock {\em The Journal of Finance\/}~{\em 22\/}(1), 19--36.

\bibitem[\protect\citeauthoryear{Bakshi, Kapadia, and Madan}{Bakshi
  et~al.}{2003}]{bakshi2003stock}
Bakshi, G., N.~Kapadia, and D.~Madan (2003).
\newblock Stock return characteristics, skew laws, and the differential pricing
  of individual equity options.
\newblock {\em The Review of Financial Studies\/}~{\em 16\/}(1), 101--143.

\bibitem[\protect\citeauthoryear{Bali, Cakici, and Whitelaw}{Bali
  et~al.}{2011}]{bali2011}
Bali, T.~G., N.~Cakici, and R.~F. Whitelaw (2011).
\newblock Maxing out: Stocks as lotteries and the cross-section of expected
  returns.
\newblock {\em Journal of Financial Economics\/}~{\em 99\/}(2), 427--446.

\bibitem[\protect\citeauthoryear{Bandi, Chaudhuri, Lo, and Tamoni}{Bandi
  et~al.}{2021}]{bandi2019spectral}
Bandi, F.~M., S.~Chaudhuri, A.~W. Lo, and A.~Tamoni (2021).
\newblock Spectral factor models.
\newblock {\em Journal of Financial Economics\/}~{\em forthcoming}.

\bibitem[\protect\citeauthoryear{Bandi and Tamoni}{Bandi and
  Tamoni}{2023}]{bandi2023}
Bandi, F.~M. and A.~Tamoni (2023).
\newblock Business-cycle consumption risk and asset prices.
\newblock {\em Journal of Econometrics\/}.

\bibitem[\protect\citeauthoryear{Bansal and Yaron}{Bansal and
  Yaron}{2004}]{bansal2004}
Bansal, R. and A.~Yaron (2004).
\newblock Risks for the long run: A potential resolution of asset pricing
  puzzles.
\newblock {\em The journal of Finance\/}~{\em 59\/}(4), 1481--1509.

\bibitem[\protect\citeauthoryear{Barberis and Huang}{Barberis and
  Huang}{2008}]{barberis2008}
Barberis, N. and M.~Huang (2008).
\newblock Stocks as lotteries: The implications of probability weighting for
  security prices.
\newblock {\em American Economic Review\/}~{\em 98\/}(5), 2066--2100.

\bibitem[\protect\citeauthoryear{Barunik and Ellington}{Barunik and
  Ellington}{2020}]{barunik2020dynamic}
Barunik, J. and M.~Ellington (2020).
\newblock Dynamic networks in large financial and economic systems.
\newblock {\em arXiv preprint arXiv:2007.07842\/}.

\bibitem[\protect\citeauthoryear{Benartzi and Thaler}{Benartzi and
  Thaler}{1995}]{benartzi1995}
Benartzi, S. and R.~H. Thaler (1995).
\newblock Myopic loss aversion and the equity premium puzzle.
\newblock {\em The quarterly journal of Economics\/}~{\em 110\/}(1), 73--92.

\bibitem[\protect\citeauthoryear{Bidder and Dew-Becker}{Bidder and
  Dew-Becker}{2016}]{bidder2016long}
Bidder, R. and I.~Dew-Becker (2016).
\newblock Long-run risk is the worst-case scenario.
\newblock {\em American Economic Review\/}~{\em 106\/}(9), 2494--2527.

\bibitem[\protect\citeauthoryear{Bollerslev, Li, and Zhao}{Bollerslev
  et~al.}{2020}]{bollerslev2018}
Bollerslev, T., S.~Z. Li, and B.~Zhao (2020).
\newblock Good volatility, bad volatility and the cross section of stock
  returns.
\newblock {\em Journal of Financial and Quantitative Analysis (JFQA)\/}~{\em
  55\/}(3), 751--781.

\bibitem[\protect\citeauthoryear{Boyer, Mitton, and Vorkink}{Boyer
  et~al.}{2009}]{boyer2009}
Boyer, B., T.~Mitton, and K.~Vorkink (2009).
\newblock Expected idiosyncratic skewness.
\newblock {\em The Review of Financial Studies\/}~{\em 23\/}(1), 169--202.

\bibitem[\protect\citeauthoryear{Campbell}{Campbell}{1996}]{campbell1996understanding}
Campbell, J.~Y. (1996).
\newblock Understanding risk and return.
\newblock {\em Journal of Political Economy\/}~{\em 104\/}(2), 298--345.

\bibitem[\protect\citeauthoryear{Chabi-Yo}{Chabi-Yo}{2012}]{chabi2012pricing}
Chabi-Yo, F. (2012).
\newblock Pricing kernels with stochastic skewness and volatility risk.
\newblock {\em Management Science\/}~{\em 58\/}(3), 624--640.

\bibitem[\protect\citeauthoryear{Chang, Christoffersen, and Jacobs}{Chang
  et~al.}{2013}]{chang2013}
Chang, B.~Y., P.~Christoffersen, and K.~Jacobs (2013).
\newblock Market skewness risk and the cross section of stock returns.
\newblock {\em Journal of Financial Economics\/}~{\em 107\/}(1), 46--68.

\bibitem[\protect\citeauthoryear{Chen}{Chen}{2002}]{chen2002intertemporal}
Chen, J. (2002).
\newblock Intertemporal capm and the cross-section of stock returns.
\newblock In {\em EFA 2002 Berlin Meetings Discussion Paper}.

\bibitem[\protect\citeauthoryear{Conrad, Dittmar, and Ghysels}{Conrad
  et~al.}{2013}]{conrad2013}
Conrad, J., R.~F. Dittmar, and E.~Ghysels (2013).
\newblock Ex ante skewness and expected stock returns.
\newblock {\em The Journal of Finance\/}~{\em 68\/}(1), 85--124.

\bibitem[\protect\citeauthoryear{Dew-Becker and Giglio}{Dew-Becker and
  Giglio}{2016}]{dew2013asset}
Dew-Becker, I. and S.~Giglio (2016).
\newblock Asset pricing in the frequency domain: theory and empirics.
\newblock {\em Review of Financial Studies\/}~{\em 29\/}(8), 2029--2068.

\bibitem[\protect\citeauthoryear{Dittmar}{Dittmar}{2002}]{dittmar2002}
Dittmar, R.~F. (2002).
\newblock Nonlinear pricing kernels, kurtosis preference, and evidence from the
  cross section of equity returns.
\newblock {\em The Journal of Finance\/}~{\em 57\/}(1), 369--403.

\bibitem[\protect\citeauthoryear{Elliott, Golub, and Jackson}{Elliott
  et~al.}{2014}]{elliott2014financial}
Elliott, M., B.~Golub, and M.~O. Jackson (2014).
\newblock Financial networks and contagion.
\newblock {\em American Economic Review\/}~{\em 104\/}(10), 3115--53.

\bibitem[\protect\citeauthoryear{Epstein and Zin}{Epstein and
  Zin}{2013}]{epstein2013substitution}
Epstein, L.~G. and S.~E. Zin (2013).
\newblock Substitution, risk aversion and the temporal behavior of consumption
  and asset returns: A theoretical framework.
\newblock In {\em Handbook of the Fundamentals of Financial Decision Making:
  Part I}, pp.\  207--239. World Scientific.

\bibitem[\protect\citeauthoryear{Fama}{Fama}{1965}]{fama1965portfolio}
Fama, E.~F. (1965).
\newblock Portfolio analysis in a stable paretian market.
\newblock {\em Management science\/}~{\em 11\/}(3), 404--419.

\bibitem[\protect\citeauthoryear{Fama}{Fama}{1976}]{fama1976}
Fama, E.~F. (1976).
\newblock {\em Foundations of finance: portfolio decisions and securities
  prices}.
\newblock Basic Books (AZ).

\bibitem[\protect\citeauthoryear{Fama}{Fama}{1996}]{fama1996}
Fama, E.~F. (1996).
\newblock Discounting under uncertainty.
\newblock {\em Journal of Business\/}, 415--428.

\bibitem[\protect\citeauthoryear{Fama and MacBeth}{Fama and
  MacBeth}{1973}]{fama1973regression}
Fama, E.~F. and J.~D. MacBeth (1973).
\newblock Risk, return, and equilibrium: Empirical tests.
\newblock {\em Journal of Political Economy\/}~{\em 81\/}(3), 607--636.

\bibitem[\protect\citeauthoryear{Farago and T{\'e}dongap}{Farago and
  T{\'e}dongap}{2018}]{farago2018downside}
Farago, A. and R.~T{\'e}dongap (2018).
\newblock Downside risks and the cross-section of asset returns.
\newblock {\em Journal of Financial Economics\/}~{\em 129\/}(1), 69--86.

\bibitem[\protect\citeauthoryear{Ghysels, Plazzi, and Valkanov}{Ghysels
  et~al.}{2016}]{ghysels2016invest}
Ghysels, E., A.~Plazzi, and R.~Valkanov (2016).
\newblock Why invest in emerging markets? the role of conditional return
  asymmetry.
\newblock {\em The Journal of Finance\/}~{\em 71\/}(5), 2145--2192.

\bibitem[\protect\citeauthoryear{Gonzalo and Olmo}{Gonzalo and
  Olmo}{2016}]{gonzalo2016}
Gonzalo, J. and J.~Olmo (2016).
\newblock Long-term optimal portfolio allocation under dynamic horizon-specific
  risk aversion.
\newblock Technical report, Universidad Carlos III de Madrid. Departamento de
  Econom{\'\i}a.

\bibitem[\protect\citeauthoryear{Gressis, Philippatos, and Hayya}{Gressis
  et~al.}{1976}]{gressis1976multiperiod}
Gressis, N., G.~C. Philippatos, and J.~Hayya (1976).
\newblock Multiperiod portfolio analysis and the inefficiency of the market
  portfolio.
\newblock {\em The Journal of Finance\/}~{\em 31\/}(4), 1115--1126.

\bibitem[\protect\citeauthoryear{Hansen and Jagannathan}{Hansen and
  Jagannathan}{1991}]{hansen1991implications}
Hansen, L.~P. and R.~Jagannathan (1991).
\newblock Implications of security market data for models of dynamic economies.
\newblock {\em Journal of Political Economy\/}~{\em 99\/}(2), 225--262.

\bibitem[\protect\citeauthoryear{Harvey, Liu, and Zhu}{Harvey
  et~al.}{2016}]{harvey2016}
Harvey, C.~R., Y.~Liu, and H.~Zhu (2016).
\newblock {\ldots} and the cross-section of expected returns.
\newblock {\em The Review of Financial Studies\/}~{\em 29\/}(1), 5--68.

\bibitem[\protect\citeauthoryear{Harvey and Siddique}{Harvey and
  Siddique}{2000}]{harvey2000}
Harvey, C.~R. and A.~Siddique (2000).
\newblock Conditional skewness in asset pricing tests.
\newblock {\em The Journal of Finance\/}~{\em 55\/}(3), 1263--1295.

\bibitem[\protect\citeauthoryear{Hou and Loh}{Hou and Loh}{2016}]{hou2016have}
Hou, K. and R.~K. Loh (2016).
\newblock Have we solved the idiosyncratic volatility puzzle?
\newblock {\em Journal of Financial Economics\/}~{\em 121\/}(1), 167--194.

\bibitem[\protect\citeauthoryear{Jondeau, Zhang, and Zhu}{Jondeau
  et~al.}{2019}]{jondeau2019average}
Jondeau, E., Q.~Zhang, and X.~Zhu (2019).
\newblock Average skewness matters.
\newblock {\em Journal of Financial Economics\/}~{\em 134\/}(1), 29--47.

\bibitem[\protect\citeauthoryear{Kamara, Korajczyk, Lou, and Sadka}{Kamara
  et~al.}{2016}]{kamara2016horizon}
Kamara, A., R.~A. Korajczyk, X.~Lou, and R.~Sadka (2016).
\newblock Horizon pricing.
\newblock {\em Journal of Financial and Quantitative Analysis\/}~{\em 51\/}(6),
  1769--1793.

\bibitem[\protect\citeauthoryear{Kelly and Jiang}{Kelly and
  Jiang}{2014}]{kelly2014tail}
Kelly, B. and H.~Jiang (2014).
\newblock Tail risk and asset prices.
\newblock {\em The Review of Financial Studies\/}~{\em 27\/}(10), 2841--2871.

\bibitem[\protect\citeauthoryear{Kim and Nelson}{Kim and
  Nelson}{2013}]{kim2013}
Kim, Y. and C.~R. Nelson (2013).
\newblock Pricing stock market volatility: does it matter whether the
  volatility is related to the business cycle?
\newblock {\em Journal of Financial Econometrics\/}~{\em 12\/}(2), 307--328.

\bibitem[\protect\citeauthoryear{Kimball}{Kimball}{1993}]{kimball1993standard}
Kimball, M.~S. (1993).
\newblock Standard risk aversion.
\newblock {\em Econometrica: Journal of the Econometric Society\/}, 589--611.

\bibitem[\protect\citeauthoryear{Kraus and Litzenberger}{Kraus and
  Litzenberger}{1976}]{kraus1976}
Kraus, A. and R.~H. Litzenberger (1976).
\newblock Skewness preference and the valuation of risk assets.
\newblock {\em The Journal of Finance\/}~{\em 31\/}(4), 1085--1100.

\bibitem[\protect\citeauthoryear{Langlois}{Langlois}{2020}]{langlois2020measuring}
Langlois, H. (2020).
\newblock Measuring skewness premia.
\newblock {\em Journal of Financial Economics\/}~{\em 135\/}(2), 399--424.

\bibitem[\protect\citeauthoryear{Lee, Wu, and Wei}{Lee
  et~al.}{1990}]{lee1990heterogeneous}
Lee, C.~F., C.~Wu, and K.~J. Wei (1990).
\newblock The heterogeneous investment horizon and the capital asset pricing
  model: theory and implications.
\newblock {\em Journal of Financial and Quantitative Analysis\/}~{\em 25\/}(3),
  361--376.

\bibitem[\protect\citeauthoryear{Lettau, Maggiori, and Weber}{Lettau
  et~al.}{2014}]{lettau2014conditional}
Lettau, M., M.~Maggiori, and M.~Weber (2014).
\newblock Conditional risk premia in currency markets and other asset classes.
\newblock {\em Journal of Financial Economics\/}~{\em 114\/}(2), 197--225.

\bibitem[\protect\citeauthoryear{Levhari and Levy}{Levhari and
  Levy}{1977}]{levhari1977capital}
Levhari, D. and H.~Levy (1977).
\newblock The capital asset pricing model and the investment horizon.
\newblock {\em The Review of Economics and Statistics\/}, 92--104.

\bibitem[\protect\citeauthoryear{Levy}{Levy}{1972}]{levy1972portfolio}
Levy, H. (1972).
\newblock Portfolio performance and the investment horizon.
\newblock {\em Management Science\/}~{\em 18\/}(12), B--645.

\bibitem[\protect\citeauthoryear{Li and Zhang}{Li and
  Zhang}{2016}]{li2016short}
Li, J. and H.~H. Zhang (2016).
\newblock Short-run and long-run consumption risks, dividend processes, and
  asset returns.
\newblock {\em The Review of Financial Studies\/}~{\em 30\/}(2), 588--630.

\bibitem[\protect\citeauthoryear{Maheu}{Maheu}{2005}]{maheu2005can}
Maheu, J. (2005).
\newblock Can garch models capture long-range dependence?
\newblock {\em Studies in Nonlinear Dynamics and Econometrics\/}~{\em 9\/}(4).

\bibitem[\protect\citeauthoryear{Maheu and McCurdy}{Maheu and
  McCurdy}{2007}]{maheu2007}
Maheu, J.~M. and T.~H. McCurdy (2007).
\newblock Components of market risk and return.
\newblock {\em Journal of Financial Econometrics\/}~{\em 5\/}(4), 560--590.

\bibitem[\protect\citeauthoryear{Maheu, McCurdy, and Zhao}{Maheu
  et~al.}{2013}]{maheu2013}
Maheu, J.~M., T.~H. McCurdy, and X.~Zhao (2013).
\newblock Do jumps contribute to the dynamics of the equity premium?
\newblock {\em Journal of Financial Economics\/}~{\em 110\/}(2), 457--477.

\bibitem[\protect\citeauthoryear{McLean and Pontiff}{McLean and
  Pontiff}{2016}]{mclean2016}
McLean, R.~D. and J.~Pontiff (2016).
\newblock Does academic research destroy stock return predictability?
\newblock {\em The Journal of Finance\/}~{\em 71\/}(1), 5--32.

\bibitem[\protect\citeauthoryear{Mehra and Prescott}{Mehra and
  Prescott}{1985}]{mehra1985equity}
Mehra, R. and E.~C. Prescott (1985).
\newblock The equity premium: A puzzle.
\newblock {\em Journal of monetary Economics\/}~{\em 15\/}(2), 145--161.

\bibitem[\protect\citeauthoryear{Merton}{Merton}{1973}]{merton1973intertemporal}
Merton, R.~C. (1973).
\newblock An intertemporal capital asset pricing model.
\newblock {\em Econometrica: Journal of the Econometric Society\/}, 867--887.

\bibitem[\protect\citeauthoryear{Mitton and Vorkink}{Mitton and
  Vorkink}{2007}]{mitton2007}
Mitton, T. and K.~Vorkink (2007).
\newblock Equilibrium underdiversification and the preference for skewness.
\newblock {\em The Review of Financial Studies\/}~{\em 20\/}(4), 1255--1288.

\bibitem[\protect\citeauthoryear{Neuberger and Payne}{Neuberger and
  Payne}{2021}]{neuberger2019skewness}
Neuberger, A. and R.~Payne (2021).
\newblock The skewness of the stock market over long horizons.
\newblock {\em The Review of Financial Studies\/}~{\em 34\/}(3), 1572--1616.

\bibitem[\protect\citeauthoryear{Neuhierl and Varneskov}{Neuhierl and
  Varneskov}{2021}]{neuhierl2021frequency}
Neuhierl, A. and R.~T. Varneskov (2021).
\newblock Frequency dependent risk.
\newblock {\em Journal of Financial Economics\/}~{\em 140\/}(2), 644--675.

\bibitem[\protect\citeauthoryear{Ortu, Tamoni, and Tebaldi}{Ortu
  et~al.}{2013}]{ortu2013}
Ortu, F., A.~Tamoni, and C.~Tebaldi (2013).
\newblock Long-run risk and the persistence of consumption shocks.
\newblock {\em Review of Financial Studies\/}~{\em 26\/}(11), 2876--2915.

\bibitem[\protect\citeauthoryear{Simkowitz and Beedles}{Simkowitz and
  Beedles}{1978}]{simkowitz1978}
Simkowitz, M.~A. and W.~L. Beedles (1978).
\newblock Diversification in a three-moment world.
\newblock {\em Journal of Financial and Quantitative Analysis\/}~{\em 13\/}(5),
  927--941.

\bibitem[\protect\citeauthoryear{Tobin}{Tobin}{1965}]{tobin1965money}
Tobin, J. (1965).
\newblock Money and economic growth.
\newblock {\em Econometrica: Journal of the Econometric Society\/}, 671--684.

\bibitem[\protect\citeauthoryear{Yu}{Yu}{2012}]{yu2012using}
Yu, J. (2012).
\newblock Using long-run consumption-return correlations to test asset pricing
  models.
\newblock {\em Review of Economic Dynamics\/}~{\em 15\/}(3), 317--335.

\end{thebibliography}
\endgroup

\newpage
\setcounter{section}{0}
\setcounter{equation}{0}
\setcounter{figure}{0}
\setcounter{table}{0}

\def\thesection{\Alph{section}}
\def\thesubsection{\thesection.\arabic{subsection}}
\def\thesubsubsection{\thesubsection.\arabic{subsubsection}}
\renewcommand{\theequation}{\Alph{section}.\arabic{equation}}
\renewcommand{\thetable}{A\arabic{table}}
\renewcommand{\thefigure}{A\arabic{figure}}

\begin{center}
	\Large \textbf{Appendix for} 
\end{center}
\begin{center}
	\Large
	``Frequency-Dependent Higher Moment Risks''
\end{center}

\vspace{10pt}

\begin{abstract}
This appendix presents supplementary details not included in the main body of the paper.
\end{abstract}

\newpage
\appendixwithtoc

\newpage

\section{Model derivation}
\label{app:technical}

This Section shows how to derive Equation~(\ref{eq:estimation1}) starting from the Euler equation. \cite{hansen1991implications} show that solution to the portfolio choice problem can be expressed in terms of Euler equation \citep{dittmar2002}

\begin{equation}
E_{t}(R_{t+1, i} M_{t+1} \mid \Omega_{t})=1,
\label{eq:euler}
\end{equation}
where $R_{t+1, i}$ is return of asset $i$, and $M_{t+1}$ is the pricing kernel. 

We build on the approach of \cite{maheu2013} and \cite{chabi2012pricing}, and derive the pricing kernel $M_{t+1}$ without explicitly assuming any form of utility function. We denote aggregate wealth at time $t$ by $W_t$, and we take a Taylor expansion of an unspecified utility function $U(W_{t+1})$ up to the fourth order.\footnote{Choice of $N = 4$  is justified in \cite{dittmar2002}.} Aggregate wealth in time $t+1$ is determined in a typical manner as $W_{t+1} = W_{t} (1+R_{t+1}^{w})$, where $R_{t+1}^{w}$ is the net return on aggregate wealth. $U(W_{t+1})$ is expanded around $W_{t}(1+C_{t})$, where $C_{t}$ is an arbitrary return

\begin{eqnarray}
U(W_{t+1}) & \approx & \sum_{n=0}^{4} \frac{U^{(n)}(W_{t}(1+C_{t}))}{n!}(W_{t+1}-W_{t}(1+C_{t}))^n \nonumber \\
& =  & \sum_{n=0}^{4} \frac{U^{(n)}(W_{t}(1+C_{t}))}{n!}(W_{t}(R_{t+1}^{w}-C_{t}))^n.
\label{eq:expansion}
\end{eqnarray}
Without loss of generality we can assume the initial wealth is equal to 1. Taking a derivative of the sum in Equation~(\ref{eq:expansion}) yields
\begin{equation}
U'(W_{t+1}) \approx \sum_{n=0}^{3} \frac{U^{(n+1)}(1+C_{t})}{n!}(R_{t+1}^{w}-C_{t})^n.
\label{eq:derivative}
\end{equation}
Note that we can interpret $U'(W_{t+1})$ as the marginal utility of wealth at time $t+1$. The pricing kernel represents investors' discounting between subsequent periods, it corresponds to changes in marginal utility given the time period in which the wealth is received. The pricing kernel, $M_{t+1} \equiv U'(W_{t+1})/U'(W_{t})$, can be approximated as

\begin{eqnarray}
M_{t+1} & \approx & \sum_{n=0}^{3} \frac{U^{(n+1)}(1+C_{t})}{U'(1) n!}(R_{t+1}^{w}-C_{t})^n \nonumber \\
& = & g_{0,t + 1} + g_{1,t + 1} (R_{t+1}^{w}-C_{t}) +g_{2,t + 1} (R_{t+1}^{w}-C_{t})^2 + g_{3,t + 1} (R_{t+1}^{w}-C_{t})^3,
\label{eq:kernel}
\end{eqnarray}
where $g_{n,t + 1} = [U^{(n+1)}(1+C_{t})/U'(1)][1 / n!] = [U^{(n+1)}(1+C_{t}) / U'(1+C_{t}) n!] [U'(1+C_{t}) / U'(1)]$. 

When we assume that investor decides between a pool of risky assets which yields the return on aggregate wealth $R_{t+1}^{w}$, and the risk-free asset yielding $R_t^f$, solution to the portfolio choice can be expressed as 
\begin{equation}
E_{t}(R_{t+1}^{w} M_{t+1} \mid \Omega_{t})=1. 
\label{eq:euler_rw}
\end{equation}
Using the formula $cov(X, Y) = E(XY) - E(X)E(Y)$, i.e. $E(XY) = cov(X, Y) + E(X)E(Y)$, we can rearrange Equation~(\ref{eq:euler_rw}) as
\begin{equation}
\begin{split}
cov_t(R_{t+1}^w, M_{t+1}) + E_t (R_{t+1}^w) E_t(M_{t+1}) = 1, \\
E_t (R_{t+1}^w) E_t(M_{t+1}) = 1 - cov_t(R_{t+1}^w, M_{t+1}), \\
E_t (R_{t+1}^w) = \frac{1}{E_t(M_{t+1})} - \frac{cov_t(R_{t+1}^w, M_{t+1})}{E_t(M_{t+1})}.
\end{split}
\label{eq:cov_dec}
\end{equation}
Recall that $1/E_t (M_{t+1}) = R_t^f$, then Equation~(\ref{eq:cov_dec}) becomes
\begin{equation}
\label{eq:cov}
E_t (R_{t+1}^w) - R_t^f = -R_t^f cov_t(R_{t+1}^w, M_{t+1}).
\end{equation}
Substituting the pricing kernel $M_{t+1}$ from Equation~(\ref{eq:kernel}) into Equation~(\ref{eq:cov}) we obtain
\begin{eqnarray}
E_{t}\left(R_{t+1}^{w}\right) - R_{t}^{f} & = & \theta_{1,t + 1} cov_t \left[R_{t+1}^{w}, R_{t+1}^{w}-C_{t}\right] + \theta_{2,t + 1} cov_t \left[R_{t+1}^{w}, \left(R_{t+1}^{w}-C_{t} \right)^2 \right] + \nonumber \\
& & \theta_{3,t + 1} cov_t \left[R_{t+1}^{w}, \left(R_{t+1}^{w}-C_{t} \right)^3 \right], 
\label{eq:rp}
\end{eqnarray}
where $R_{t}^{f}$ is the risk-free rate, and $\theta_{n,t + 1} = -g_{n,t + 1}R_{t}^{f}$. 

At this point, it is convenient to specify the expansion point. A widely used choice is $C_{t}=0$ \citep[e.g.,][]{harvey2000, dittmar2002}, an alternative approach is to set $C_{t}=E_{t}(R_{t+1}^{w})$. We employ the latter specification also used by \cite{chabi2012pricing} or \cite{maheu2013}. \cite{chabi2012pricing} shows that this approach is equivalent to small noise expansion, if we write
\begin{equation}
R_{t+1}^{w} - E_{t} (R_{t+1}^{w}) = \epsilon Y_{t+1},
\label{eq:error}
\end{equation}
then driving $\epsilon$ towards zero causes $R_{t+1}^{w}$ to approach $E_{t} (R_{t+1}^{w})$. Let us simplify notation by denoting for each $i \in \{ 1, \dots, N \}$

\begin{eqnarray}
R_{t+1, i}^{e} = R_{t+1, i} - R_{t}^{f}, \\
R_{t+1} = R_{t+1}^{w} - R_{t}^{f}.
\end{eqnarray} 
Next, we can rewrite Equation~({\ref{eq:error}) in terms of excess returns on aggregate wealth
\begin{equation}
\epsilon_{t+1} = R_{t+1}^{w} - E_{t} (R_{t+1}^{w}) = (R_{t+1}^{w} - R_{t}^{f}) - (E_{t} (R_{t+1}^{w}) - R_{t}^{f}) = R_{t+1} - E_{t} (R_{t+1}).
\label{eq:epsilon}
\end{equation}

\cite{jondeau2019average} suggest that the return of asset $i$ can be decomposed to the systemic component and purely idiosyncratic component, thus for each $i$ we can write $R_{t+1, i}^{e}$ in a following way
\begin{equation}
R_{t+1, i}^{e} = R_{t + 1}^{(m)} + R_{t + 1}^{(i)},
\label{eq:risk}
\end{equation}
where $R_{t + 1}^{(m)}$ is the systemic (market) component, and $R_{t + 1}^{(i)}$ is the idiosyncratic component of the return of asset $i$. Assuming that the market return is the same for each $i$, we express the return on aggregate wealth by aggregating for all $i \in \{1, \dots, N \}$\footnote{For details, see \cite{jondeau2019average}}
\begin{equation}
R_{t + 1} = R_{t + 1}^{(m)} + \sum_{i = 1}^{N} w_i R_{t + 1}^{(i)},
\label{eq:rw_risk}
\end{equation}
where $N$ is the number of assets. We assign an equal weight to each asset, hence we can write
\begin{equation}
R_{t + 1} = R_{t + 1}^{(m)} + \frac{1}{N} \sum_{i = 1}^{N} R_{t + 1}^{(i)}.
\label{eq:rw_risk_equal}
\end{equation}
Based on Equation~(\ref{eq:rw_risk_equal}), $\epsilon_{t + 1}$ can also be decomposed to the systemic and idiosyncratic part
\begin{equation}
\epsilon_{t+1} = \left[R_{t+1}^{(m)} - E_{t} \left(R_{t+1}^{(m)} \right) \right] + \frac{1}{N} \sum_{i = 1}^{N} \left[R_{t+1}^{(i)} - E_{t} \left(R_{t+1}^{(i)} \right) \right] = \epsilon_{t+1}^{(m)} + \frac{1}{N} \sum_{i = 1}^{N} \epsilon_{t+1}^{(i)}.
\label{eq:eps_risk}
\end{equation}
In addition to decomposing returns on aggregate wealth to the systemic and idiosyncratic part, these can also be decomposed to components representing different investment horizons as we show in Section~\ref{subsec:model}. Specifically, we decompose returns to the transitory and persistent components
\begin{equation}
R_{t+1} \equiv \sum_{j=1}^{N} R_{t+1}^{(j)} + R_{t+1}^{(\infty)} = R_{t+1}^{(\tr)} + R_{t+1}^{(\p)},
\label{eq:ret_dec}
\end{equation}
where $R_{t+1}^{(\tr)} = \sum_{j=1}^{N} R_{t+1}^{(j)}$ is the transitory component of returns of aggregate wealth, $R_{t+1}^{(\p)} = R_{t+1}^{(\infty)} = R_{t+1}^{(n > N)}$ is the persistent component of returns of aggregate wealth.\footnote{Note that due to the equivalence in Equation~(\ref{eq:ret_dec}), the decomposition is not restricted to two horizons. In fact, we are able to construct components from arbitrary number of horizons by splitting the sum in intermediate points.} 

Equation~(\ref{eq:rw_risk_equal}) shows that returns on aggregate wealth are represented by aggregating the systemic component $R_{t + 1}^{(m)}$, and a weighted sum of individual purely idiosyncratic components $R_{t + 1}^{(i)}$ for $i \in \{1, \dots, N \}$. We can apply such decomposition to both of the horizon-specific components of returns on aggregate wealth from Equation~(\ref{eq:ret_dec}), and rewrite them as
\begin{equation}
\begin{split}
R_{t+1}^{(\tr)} & =  R_{t+1}^{(m, \tr)} + \frac{1}{N} \sum_{i = 1}^{N} R_{t+1}^{(i, \tr)}, \\
R_{t+1}^{(\p)} & =  R_{t+1}^{(m, \p)} + \frac{1}{N} \sum_{i = 1}^{N} R_{t+1}^{(i, \p)},
\end{split}
\label{eq:ret_dec_detail}
\end{equation}
where $r \in \{ m, i \}$ denotes systemic ($m$) risk, and idiosyncratic ($i$) risk, $h \in \{ s, l \}$ denotes the transitory ($\tr$) and the persistent ($\p$) component of returns. By combining Equation~(\ref{eq:ret_dec_detail}) and Equation~(\ref{eq:ret_dec}), we obtain returns on aggregate wealth consisting of four components accommodating the decomposition to market and idiosyncratic components as well as the decomposition to transitory and persistent components
\begin{equation}
R_{t + 1} = R_{t + 1}^{(m, \tr)} + R_{t + 1}^{(m, \p)} + \frac{1}{N} \sum_{i = 1}^{N} R_{t + 1}^{(i, \tr)} + \frac{1}{N} \sum_{i = 1}^{N} R_{t + 1}^{(i, \p)}.
\label{eq:rw_dec_final}
\end{equation}
Analogous representation of $\epsilon_{t + 1}$ directly follows from Equation~(\ref{eq:rw_dec_final})
\begin{eqnarray}
\epsilon_{t+1} & = & R_{t+1} - E_{t} (R_{t+1}) \nonumber \\ 
 & = &  \left[R_{t+1}^{(m, \tr)} - E_{t} \left(R_{t+1}^{(m, \tr)} \right) \right] +  \left[R_{t+1}^{(m, \p)} - E_{t} \left(R_{t+1}^{(m, \p)} \right) \right] +  \nonumber \\ 
& &  \frac{1}{N} \sum_{i = 1}^{N} \left[R_{t+1}^{(i, \tr)} - E_{t} \left(R_{t+1}^{(i, \tr)} \right) \right] + \frac{1}{N} \sum_{i = 1}^{N} \left[R_{t+1}^{(i, \p)} - E_{t} \left(R_{t+1}^{(i, \p)} \right) \right] \nonumber \\
& = & \epsilon_{t+1}^{(m, \tr)} + \epsilon_{t+1}^{(m, \p)} + \frac{1}{N} \sum_{i = 1}^{N} \epsilon_{t+1}^{(i, \tr)} + \frac{1}{N} \sum_{i = 1}^{N} \epsilon_{t+1}^{(i, \p)}.
\label{eq:eps_final}
\end{eqnarray}
Let us assume that for each $i$, investors invest their whole wealth into asset $i$, hence $R_{t+1, i}^{e}$ and $R_{t+1}$ can be treated as interchangeable below. This allows us to express risk premium of asset $i$ as
\begin{equation}
E_{t} \left(R_{t+1, i}^{e}\right) =  \theta_{1,t + 1, i} cov_{t} \left(R_{t+1}, \epsilon_{t+1} \right) + \theta_{2,t + 1, i} cov_{t} \left(R_{t+1}, \epsilon_{t+1}^2 \right) +
\theta_{3,t + 1, i} cov_{t} \left(R_{t+1}, \epsilon_{t+1}^3\right).
\label{eq:risk_prem}
\end{equation}

As we show in  Equation~(\ref{eq:rw_dec_final}) and Equation~(\ref{eq:eps_final}), we can decompose $R_{t + 1}$ and $\epsilon_{t + 1}$ into the four components corresponding to combinations of specific horizons and risk sources. We express Equation~(\ref{eq:risk_prem}) in terms of these components\footnote{The orthogonality assumptions allowing the transition from Equation~(\ref{eq:risk_prem}) to Equation~(\ref{eq:rt_dec}) are discussed in Appendix~\ref{sec:orthogonality}.}
\begin{eqnarray}
E_{t} \left(R_{t+1, i}^{e} \right) & = & \sum_{h \in\{\tr, \p \}}  \sum_{k = 1}^{3} \theta_{2, t + 1, i}^{(m, h)} cov_t \left( R_{t + 1}^{(m, h)}, \left( \epsilon_{t + 1}^{(m, h)} \right)^{k} \right) \nonumber \\ 
& + & \sum_{h \in\{\tr, \p \}} \sum_{k = 1}^{3} \theta_{2, t + 1, i}^{(I, h)} cov_t \left(\frac{1}{N} \sum_{i = 1}^{N} R_{t + 1}^{(i, h)}, \left( \frac{1}{N} \sum_{j = 1}^{N} \epsilon_{t + 1}^{(i, h)} \right)^k \right).
\label{eq:rt_dec}
\end{eqnarray}
To simplify the notation we aggregate the individual idiosyncratic returns, and we denote
\begin{equation*}
\begin{split}
& \frac{1}{N} \sum_{i = 1}^{N} R_{t + 1}^{(i, h)} =  R_{t + 1}^{(I, h)}, \\
& \frac{1}{N} \sum_{i = 1}^{N} \epsilon_{t + 1}^{(i, h)} =  \epsilon_{t + 1}^{(I, h)},
\end{split}
\end{equation*} 
where $ h \in\{\tr, \p \}$ denotes the corresponding horizon, and the upper index $I$ denotes the average idiosyncratic component of the returns on aggregate wealth. Then, we can rewrite Equation~(\ref{eq:rt_dec}) as
\begin{eqnarray}
E_{t} \left(R_{t+1, i}^{e} \right) & = & \sum_{h \in\{\tr, \p \}} \theta_{1, t + 1, i}^{(m, h)} cov_t \left( R_{t + 1}^{(m, h)}, \epsilon_{t + 1}^{(m, h)} \right) + \sum_{h \in\{\tr, \p \}} \theta_{1, t + 1, i}^{(I, h)} cov_t \left( R_{t + 1}^{(I, h)}, \epsilon_{t + 1}^{(I, h)} \right) \nonumber \\
& + & \sum_{h \in\{\tr, \p \}} \theta_{2, t + 1, i}^{(m, h)} cov_t \left( R_{t + 1}^{(m, h)}, \left( \epsilon_{t + 1}^{(m, h)} \right)^2 \right) + \sum_{h \in\{\tr, \p \}} \theta_{2, t + 1, i}^{(I, h)} cov_t \left( R_{t + 1}^{(I, h)}, \left( \epsilon_{t + 1}^{(I, h)} \right)^2 \right) \nonumber \\
& + & \sum_{h \in\{\tr, \p \}} \theta_{3, t + 1, i}^{(m, h)} cov_t \left( R_{t + 1}^{(m, h)}, \left( \epsilon_{t + 1}^{(m, h)} \right)^3 \right) + \sum_{h \in\{\tr, \p \}} \theta_{3, t + 1, i}^{(I, h)} cov_t \left( R_{t + 1}^{(I, h)}, \left( \epsilon_{t + 1}^{(I, h)} \right)^3 \right) \nonumber \\
& = & \sum_{ h \in\{\tr, \p \}} \sum_{ r \in \{ m, I \}} \sum_{k = 1}^{3} \theta_{k, t + 1, i}^{(r, h)} cov_t \left( R_{t + 1}^{(r, h)}, \left( \epsilon_{t + 1}^{(r, h)} \right)^k \right),
\label{eq:rt_dec_final}
\end{eqnarray}
where $ r \in \{ m, I \}$ denotes the systemic and average idiosyncratic risk respectively.
Recall that $R_{t+1}^{(r, h)} - E_{t}\left(R_{t+1}^{(r, h)}\right) = \epsilon_{t+1}^{(r, h)}$, hence for $r \in \{m, I \}$, and $h \in\{\tr, \p \}$, we can rewrite $cov_{t}\left(R_{t+1}^{(r, h)}, \epsilon_{t+1}^{(r, h)}\right)$ in the following way
\begin{eqnarray}
cov_{t}\left[R_{t+1}^{(r, h)}, \epsilon_{t+1}^{(r, h)}\right] & = & E_{t}\left[R_{t+1}^{(r, h)} \epsilon_{t+1}^{(r, h)}\right] - E_{t}\left[R_{t+1}^{(r, h)}\right] E_{t}\left[\epsilon_{t+1}^{(r, h)}\right] \nonumber \\
& = & E_{t}\left[ \left(\epsilon_{t+1}^{(r, h)} + E_{t}\left[R_{t+1}^{(r, h)}\right]\right) \epsilon_{t+1}^{(r, h)}\right] - E_{t}\left[R_{t+1}^{(r, h)}\right] E_{t}\left[\epsilon_{t+1}^{(r, h)}\right] \nonumber \\
& = & E_{t}\left[\left(\epsilon_{t+1}^{(r, h)}\right)^2 \right] + E_{t} \left[\epsilon_{t+1}^{(r, h)} E_{t}\left(R_{t+1}^{(r, h)}\right)\right] - E_{t}\left[R_{t+1}^{(r, h)}\right] E_{t}\left[\epsilon_{t+1}^{(r, h)}\right] \nonumber \\
& = & E_{t}\left[\left(\epsilon_{t+1}^{(r, h)}\right)^2 \right] = var_{t}\left[\epsilon_{t+1}E_{t}\left[\left(\epsilon_{t+1}^{(r, h)}\right)^2 \right]\right],
\label{eq:var_rh}
\end{eqnarray}
since $E_{t}\left[\epsilon_{t+1}^{(r, h)} E_{t}\left(R_{t+1}^{(r, h)}\right)\right] = E_{t}\left(R_{t+1}^{(r, h)}\right) E_{t}\left(\epsilon_{t+1}^{(r, h)}\right)$, and $E_{t}\left(\epsilon_{t+1}^{(r, h)}\right) = 0$. Analogously, we can derive that

\begin{equation}
\begin{split}
cov_{t}\left( R_{t+1}^{(r, h)}, \left(\epsilon_{t+1}^{(r, h)}\right)^2 \right) = E_{t}\left( \left(\epsilon_{t+1}^{(r, h)}\right)^3 \right), \\
cov_{t}\left( R_{t+1}^{(r, h)}, \left(\epsilon_{t+1}^{(r, h)}\right)^3 \right) = E_{t}\left( \left(\epsilon_{t+1}^{(r, h)}\right)^4 \right),
\end{split}
\label{eq:skewkurt}
\end{equation}
for $r \in \{m, I \}$, and $h \in\{\tr, \p \}$. Given the results from Equation~(\ref{eq:var_rh}) and Equation~(\ref{eq:skewkurt}), we can express  Equation~(\ref{eq:rt_dec}) as

\begin{eqnarray}
E_{t} (R_{t+1, i}^{e}) & = &  \sum_{r \in \{ m, I \}} \sum_{h \in\{\tr, \p \}}\theta_{1,t + 1, i}^{(r, h)} E_{t} \left[ \left(\epsilon_{t+1}^{(r, h)} \right)^2 \right] +  \sum_{r \in \{ m, I \}} \sum_{h \in\{\tr, \p \}}\ \theta_{2,t + 1, i}^{(r, h)} E_{t} \left[ \left(\epsilon_{t+1}^{(r, h)} \right)^3 \right]  \nonumber \\ 
& + & \sum_{r \in \{ m, I \}} \sum_{h \in\{\tr, \p \}}\ \theta_{3,t + 1, i}^{(r, h)} E_{t} \left[ \left(\epsilon_{t+1}^{(r, h)} \right)^4 \right],
\label{eq:rt_freq_app}
\end{eqnarray}
where $E_{t} \left[ \left(\epsilon_{t+1}^{(m, h)} \right)^n \right]$ denotes n-th centralized moment of the systemic component of returns on aggregate wealth, $E_{t} \left[ \left(\epsilon_{t+1}^{(I, h)} \right)^n \right]$ denotes n-th centralized moment of the idiosyncratic component of returns on aggregate wealth, and
\begin{eqnarray*}
\theta_{1,t + 1, i}^{(h)} & = & - \frac{U^{(2)}(1+C_{t})}{U'(1+C_{t})} \frac{U'(1+C_{t})R_{t}^{f}} {U'(1)} w_{1, t + 1, i}^{(h)} \omega_{1, t + 1, i}^{(r)}, \\
\theta_{2,t + 1, i}^{(h)} & = & - \frac{U^{(3)}(1+C_{t})}{U'(1+C_{t}) 2!} \frac{U'(1+C_{t})R_{t}^{f}} {U'(1)}  w_{2,t + 1, i}^{(h)} \omega_{2, t + 1, i}^{(r)}, \\
\theta_{3,t + 1, i}^{(h)} & = & - \frac{U^{(4)}(1+C_{t})}{U'(1+C_{t}) 3!} \frac{U'(1+C_{t})R_{t}^{f}} {U'(1)}  w_{3,t + 1, i}^{(h)} \omega_{3, t + 1, i}^{(r)}, 
\end{eqnarray*}
where $h \in\{\tr, \p \}$, $r \in \{ m, I \}$, $w_{k,t + 1, i}^{(h)}$, $k \in \{ 1,2,3 \}$, $i \in \{1, \dots, N \}$ are the spectral weights, and $\omega_{k, t + 1, i}^{(r)}$, $k \in \{ 1,2,3 \}$, $i \in \{1, \dots, N \}$ are the risk source weights. For sake of clarity, we define $v_{t + 1}^{(r, h)} = var_{t} \left( \epsilon_{t+1}^{(r, h)}\right)$,  $s_{t + 1}^{(r, h)} = E_{t} \left[ \left(\epsilon_{t+1}^{(r, h)} \right)^3 \right ]/ \left( v_{t + 1}^{(r, h)} \right)^{3/2}$, $k_{t + 1}^{(r, h)} = E_{t} \left[ \left(\epsilon_{t+1}^{(r, h)} \right)^4 \right]/ \left(v_{t + 1}^{(r, h)} \right)^2$. Then, the final representation of Equation~(\ref{eq:rt_freq_app}) is
\begin{equation}
E_{t} (R_{t+1, i}^{e}) =  \sum_{r \in \{ m, I \}} \sum_{h\in\{s,l\}} \beta_{t + 1, i}^{(r, h)} \sqrt{v_{t + 1}^{(r, h)}} +  \sum_{r \in \{ m, I \}} \sum_{h\in\{s,l\}} \delta_{t + 1, i}^{(r, h)} s_{t + 1}^{(r, h)} +  \sum_{r \in \{ m, I \}} \sum_{h\in\{s,l\}} \mathcal{K}_{t + 1, i}^{(r, h)} k_{t + 1}^{(r, h)},
\label{eq:rt_final_app}
\end{equation}
where $h \in\{\tr, \p \}$ and $r \in \{ m, I \}$, $\sqrt{v_{t + 1}^{(r, h)}}$, $s_{t + 1}^{(r, h)}$, and $k_{t + 1}^{(r, h)}$ denote volatility, skewness, and kurtosis of returns on aggregate wealth respectively. $x_{t + 1}^{(m, h)}$, $x \in \{v, s, k \}$, $h \in\{\tr, \p \}$, denotes the corresponding moment of the systemic component of the returns on aggregate wealth, $x_{t + 1}^{(I, h)}$, $x \in \{v, s, k \}$, $h \in\{\tr, \p \}$, denotes the corresponding moment of the average idiosyncratic component of the returns on aggregate wealth. Finally, $x_{t + 1}^{(r, \tr)}$, $x \in \{v, s, k \}$, $r \in \{ m, I \}$, denotes the transitory component of the corresponding moment, $x_{t + 1}^{(r, \p)}$, $x \in \{v, s, k \}$, $r \in \{ m, I \}$, denotes the persistent component of the corresponding moment, and 

\begin{eqnarray*}
\beta_{t + 1, i}^{(r, h)} & = & \theta_{1,t + 1, i}^{(r, h)} \left(v_{t + 1}^{(r, h)} \right)^{1/2}, \\
\delta_{t + 1, i}^{(r, h)} & = & \theta_{2,t + 1, i}^{(r, h)} \left(v_{t + 1}^{(r, h)} \right)^{3/2}, \\
\mathcal{K}_{t + 1, i}^{(r, h)} & = &  \theta_{3,t + 1, i}^{(r, h)} \left(v_{t + 1}^{(r, h)}\right)^2, \\
\end{eqnarray*}
for $h \in\{\tr, \p \}$, and $r \in \{ m, I \}$. 

To finalise, when we substitute $MMT_t \in \{ v_t, s_t, k_t \}$ by the realized higher moments $RDM_t \in \{ RDVOL_t, RDS_t, RDK_t \} $ employed in our empirical exercise, we can estimate Equation~(\ref{eq:rt_final_app}) for each $i$ using the following cross-sectional regression
\begin{eqnarray}
r_{t, i} & =  & \sum_{r \in \{ m, I \}} \sum_{h \in\{\tr, \p \}} \beta_{t, i, V}^{(r, h)} RDVOL_{t}^{(r, h)} +  \sum_{r \in \{ m, I \}} \sum_{h \in\{\tr, \p \}} \beta_{t, i, S}^{(r, h)} RDS_{t}^{(r, h)} \nonumber \\
& + & \sum_{r \in \{ m, I \}} \sum_{h \in\{\tr, \p \}} \beta_{t, i, K}^{(r, h)} RDK_{t}^{(r, h)},
\label{eq:app_fm1}
\end{eqnarray}
where $RDVOL$ denoted Realized Volatility, $RDS$ denotes Realized Skewness, and $RDK$ denotes Realized Kurtosis. Finally, the sensitivities to individual risks obtained from Equation~(\ref{eq:app_fm1}) are used to predict subsequent returns of each asset $i$ using the following second-stage regression
\begin{eqnarray}
r_{t + 1, i} & = &   \sum_{r \in \{ m, I \}} \sum_{h \in\{\tr, \p \}}  \lambda_{t, V}^{(r, h)} \beta_{t, i, V}^{(r, h)} +  \sum_{r \in \{ m, I \}} \sum_{h \in\{\tr, \p \}}  \lambda_{t, S}^{(r, h)} \beta_{t, i, S}^{(r, h)}  \nonumber \\ 
& + & \sum_{r \in \{ m, I \}} \sum_{h \in\{\tr, \p \}}  \lambda_{t, K}^{(r, h)} \beta_{t, i, K}^{(r, h)}.
\label{eq:app_fm2}
\end{eqnarray}

\newpage
\section{Simulations}
\label{sec:simulation}

The properties of $r_{t, n}^{(\tr)}$ and $r_{t, n}^{(\p)}$ in Equation~(\ref{eq:ret_simulation}) depend on the evolution of the transitory and persistent variance and skewness processes formalized below. The overall variance of the continuous time returns process is composed of the transitory and the persistent variance components
\begin{eqnarray}
\sigma_t^2 &=& \left( \sigma_t^{(\tr)} +\sigma_t^{(\p)} \right)^2,  \\
d\left(\sigma_t^{(\tr)} \right)^2 &=& \theta_1 \left( \theta_2 - \left(\sigma_t^{(\tr)} \right)^2 \right) dt + \theta_3 \left(\sigma_t^{(\tr)} \right)^2 dW_1(t), \\ 
d\left(\sigma_t^{(\p)} \right)^2 &=& \rho_1 \left( \rho_2 - \left(\sigma_t^{(\p)} \right)^2 \right) dt + \rho_3 \left(\sigma_t^{(\p)} \right)^2 dW_2(t),
\label{eq:var_simulation}
\end{eqnarray}
where $W_1(t)$ and $W_2(t)$ are Brownian motions. The skewness parameters $\lambda_{t-1}^{(s)}$ and $\lambda_{t-1}^{(l)}$ are obtained from the skewness processes
\begin{eqnarray}
d\gamma_t^{(s)} &=& \phi_1 \left( \phi_2 - \gamma_t^{(s)} \right) dt + \phi_3 dW_3(t) + \phi_4 \gamma_t^{(s)} dW_3^3(t),  \\ 
\lambda_t^{(s)} &=& -1 + \frac{2}{1 + exp \left( -\gamma_t^{(s)} \right) }, \\
d\gamma_t^{(l)} &=& \kappa_1 \left( \kappa_2 - \gamma_t^{(l)} \right) dt + \kappa_3  dW_4(t), \\
\lambda_t^{(l)} &=& -1 + \frac{2}{1 + exp \left( -\gamma_t^{(l)} \right) }, 
\label{eq:skew_simulation}
\end{eqnarray}
where $\gamma_t^{(s)}$ is the transitory skewness process, and $\gamma_t^{(l)}$ is the persistent skewness process, $W_3(t)$ and $W_4(t)$ are Brownian motions.

\newpage
\section{Orthogonality}
\label{sec:orthogonality}

After expressing Equation~(\ref{eq:risk_prem}) in terms of the individual components of $R_{t + 1}$ and $\epsilon_{t + 1}$, we in fact obtain more covariance terms than are displayed in Equation~(\ref{eq:rt_dec}). We can eliminate number of covariance terms due to orthogonality of the individual components described below. The components corresponding to different scales are orthogonal by construction of the frequency decomposition, i.e. components representing different horizons are orthogonal
\begin{eqnarray*}
R_{t + 1}^{(m, \tr)} \epsilon_{t + 1}^{(m, \p)} = R_{t + 1}^{(m, \p)} \epsilon_{t + 1}^{(m, \tr)} & = & 0, \\
\epsilon_{t + 1}^{(m, \tr)} \epsilon_{t + 1}^{(m, \p)} & = & 0, \\
R_{t + 1}^{(i, \tr)} \epsilon_{t + 1}^{(j, \p)} = R_{t + 1}^{(i, \p)} \epsilon_{t + 1}^{(j, \tr)} & = &  0,\\ 
\epsilon_{t + 1}^{(i, \tr)} \epsilon_{t + 1}^{(j, \p)} & = & 0,
\end{eqnarray*}
for all combinations of $i \in \{ 1, \dots, N \}$, and $j \in \{ 1, \dots, N \}$.

\cite{jondeau2019average} implicitly assume that the systemic components, and the individual idiosyncratic components are orthogonal, i.e.
\begin{eqnarray*}
R_{t + 1}^{(m, \tr)} \epsilon_{t + 1}^{(i, \tr)} = R_{t + 1}^{(m, \tr)} \epsilon_{t + 1}^{(i, \p)} = R_{t + 1}^{(m, \p)} \epsilon_{t + 1}^{(i, \tr)} = R_{t + 1}^{(m, \p)} \epsilon_{t + 1}^{(i, \p)} & = & 0, \\
R_{t + 1}^{(i, \tr)} \epsilon_{t + 1}^{(m, \tr)} = R_{t + 1}^{(i, \tr)} \epsilon_{t + 1}^{(m, \p)} = R_{t + 1}^{(i, \p)} \epsilon_{t + 1}^{(m, \tr)} = R_{t + 1}^{(i, \p)} \epsilon_{t + 1}^{(m, \p)} & = & 0, \\ 
\epsilon_{t + 1}^{(m, \tr)} \epsilon_{t + 1}^{(i, \tr)} = \epsilon_{t + 1}^{(m, \tr)} \epsilon_{t + 1}^{(i, \p)} =  \epsilon_{t + 1}^{(m, \p)} \epsilon_{t + 1}^{(i, \tr)} =  \epsilon_{t + 1}^{(m, \p)} \epsilon_{t + 1}^{(i, \p)} & = & 0,\\
\end{eqnarray*}
for all $i \in \{ 1, \dots, N \}$.
In addition, we assume that the orthogonality holds also for the second powers of $\epsilon_{t + 1}^{(r, h)}$. Specifically, 
\begin{eqnarray*}
\left( \epsilon_{t + 1}^{(m, \tr)} \right)^2 \epsilon_{t + 1}^{(m, \p)} = \epsilon_{t + 1}^{(m, \tr)} \left( \epsilon_{t + 1}^{(m, \p)} \right)^2 & = & 0, \\
\left( \epsilon_{t + 1}^{(m, \tr)} \right)^2 \epsilon_{t + 1}^{(i, \tr)} = \left( \epsilon_{t + 1}^{(m, \tr)} \right)^2 \epsilon_{t + 1}^{(i, \p)} =  \left( \epsilon_{t + 1}^{(m, \p)} \right)^2 \epsilon_{t + 1}^{(i, \tr)} = \left( \epsilon_{t + 1}^{(m, \p)} \right)^2 \epsilon_{t + 1}^{(i, \p)} & = & 0, \\ 
\left( \epsilon_{t + 1}^{(i, \tr)} \right)^2 \epsilon_{t + 1}^{(m, \tr)} = \left( \epsilon_{t + 1}^{(i, \tr)} \right)^2 \epsilon_{t + 1}^{(m, \p)} =  \left( \epsilon_{t + 1}^{(i, \p)} \right)^2 \epsilon_{t + 1}^{(m, \tr)} = \left( \epsilon_{t + 1}^{(i, \p)} \right)^2 \epsilon_{t + 1}^{(m, \p)} & = & 0, \\ 
\end{eqnarray*}
for all $i \in \{ 1, \dots, N \}$, and
\begin{equation*}
\left( \epsilon_{t + 1}^{(i, \tr)} \right)^2 \epsilon_{t + 1}^{(j, \p)} = \epsilon_{t + 1}^{(i, \tr)} \left( \epsilon_{t + 1}^{(j, \p)} \right)^2 = 0, 
\end{equation*}
for all combinations of $i \in \{ 1, \dots, N \}$, and $j \in \{ 1, \dots, N \}$. Lastly, we assume that in some cases the orthogonality holds for the second, and third power of $\epsilon_{t + 1}^{(r, h)}$, i.e
\begin{eqnarray*}
R_{t + 1}^{(m, \tr)} \left(\epsilon_{t + 1}^{(m, \p)} \right)^k = R_{t + 1}^{(m, \p)} \left(\epsilon_{t + 1}^{(m, \tr)} \right)^k  & = & 0, \\
R_{t + 1}^{(m, \tr)} \left(\epsilon_{t + 1}^{(i, \tr)} \right)^k = R_{t + 1}^{(m, \tr)} \left(\epsilon_{t + 1}^{(i, \p)} \right)^k = R_{t + 1}^{(m, \p)} \left(\epsilon_{t + 1}^{(i, \tr)} \right)^k = R_{t + 1}^{(m, \p)} \left(\epsilon_{t + 1}^{(i, \p)} \right)^k & = & 0, \\
\left(\epsilon_{t + 1}^{(m, \tr)} \right)^k R_{t + 1}^{(i, \tr)} = \left(\epsilon_{t + 1}^{(m, \tr)} \right)^k R_{t + 1}^{(i, \p)} = \left(\epsilon_{t + 1}^{(m, \p)} \right)^k R_{t + 1}^{(i, \tr)} = \left(\epsilon_{t + 1}^{(m, \p)} \right)^k R_{t + 1}^{(i, \p)} & = & 0, \\
\end{eqnarray*}
for all $i \in \{ 1, \dots, N \}$, $k \in \{ 2, 3 \}$, and 
\begin{equation*}
\left(\epsilon_{t + 1}^{(j, \tr)} \right)^k R_{t + 1}^{(i, \p)} = \left(\epsilon_{t + 1}^{(j, \p)} \right)^k R_{t + 1}^{(i, \tr)} = 0,
\end{equation*}
for all combinations of $i \in \{ 1, \dots, N \}$, and $j \in \{ 1, \dots, N \}$, and $k \in \{ 2, 3 \}$.

\newpage
\section{Estimated models}
\label{app:models}

We emlpoy the Fama-Macbeth type cross-sectional regressions \citep{fama1973regression} where the first-stage regression is defined as
\begin{equation}
r_{t, i} = \sum_{k = 1}^K \beta_{t, i, k} X_{t, k} + \sum_{j = 1}^J \gamma_{t, i, j} Z_{t, j}
\end{equation}
with $X_{t, k}$ being the k-th higher moment variable at time t, and $Z_{t, j}$ being the j-th control variable at time t. The second-stage regression is then
\begin{equation}
r_{t + 1, i} = \sum_{k = 1}^K \lambda_{t, k} \beta_{t, i, k} + \sum_{j = 1}^J \phi_{t, j} \gamma_{t, i, j}.
\end{equation}
The individual models that we estimate differ in the content of $\mathbf{X} = \{ X_1, \dots, X_K \}$, and $\mathbf{Z} = \{ Z_1, \dots, Z_J \}$. The aim of this paper is to show that decomposing higher moments risks to the transitory and persistent components is a crucial step towards understanding how these moments are priced in the cross-section of stocks returns. Hence, we present models considering the aggregate measures of higher moment risks and the models considering heterogeneously persistent components of the higher moment risks separately. Secondly, we want to explore both the individual and joint effects of market and average idiosyncratic risk. Hence, we estimate models containing only market risk, models containing only average idiosyncratic risk, and models considering these two risk sources jointly. Lastly, the full model also contains the FF3 control variables.

Here, we present the design of the first-stage regressions for the corresponding models, we omit the second stage equations as they are implied by the first stage. The model labeled as \textbf{1} contains the market aggregate higher moments
\begin{equation}
r_{t, i} =  \beta_{t, i, V}^{(m)} RDVOL_{t}^{(m)} +  \beta_{t, i, S}^{(m)} RDS_{t}^{(m)} +  \beta_{t, i, K}^{(m)} RDK_{t}^{(m)}.
\end{equation}
The model labeled as \textbf{2} contains the average idiosyncratic aggregate higher moments
\begin{equation}
r_{t, i} =  \beta_{t, i, V}^{(I)} RDVOL_{t}^{(I)} +  \beta_{t, i, S}^{(I)} RDS_{t}^{(I)} +  \beta_{t, i, K}^{(I)} RDK_{t}^{(I)}.
\end{equation}
The model labeled as \textbf{3} combines the market and average idiosyncratic aggregate higher moments, and controls for the variables from the FF3 model
\begin{eqnarray}
r_{t, i} & =  & \sum_{r \in \{ m, I \}} \beta_{t, i, V}^{(r)} RDVOL_{t}^{(r)} +  \sum_{r \in \{ m, I \}} \beta_{t, i, S}^{(r)} RDS_{t}^{(r)}  \nonumber \\
& + & \sum_{r \in \{ m, I \}}  \beta_{t, i, K}^{(r)} RDK_{t}^{(r)} + \sum_{j = 1}^J \gamma_{t, i, j} Z_{t, j},
\end{eqnarray}
where $r \in \{ m, I \}$ indicates market and average idiosyncratic risk respectively, and $\mathbf{Z} \in \{ MKT, SMB, HML \}$ contains the FF3 control variables.

Models \textbf{4} - \textbf{6} contain the corresponding measure from models \textbf{1} - \textbf{3} decomposed to the transitory and the persistent components. Model \textbf{4} considers the heterogeneously persistent components of the market higher moments
\begin{eqnarray}
r_{t, i} & =  &  \sum_{h \in\{\tr, \p\}} \beta_{t, i, V}^{(m, h)} RDVOL_{t}^{(m, h)} +  \sum_{h \in\{\tr, \p\}} \beta_{t, i, S}^{(m, h)} RDS_{t}^{(m, h)}  \nonumber \\
& + & \sum_{h \in\{\tr, \p\}} \beta_{t, i, K}^{(m, h)} RDK_{t}^{(m, h)},
\end{eqnarray}
where $h \in\{\tr, \p\}$ denotes the transitory and persistent risk respectively. Model \textbf{5} contains the heterogeneously persistent components of the average idiosyncratic higher moments
\begin{eqnarray}
r_{t, i} & =  &  \sum_{h \in\{\tr, \p\}} \beta_{t, i, V}^{(I, h)} RDVOL_{t}^{(I, h)} +  \sum_{h \in\{\tr, \p\}} \beta_{t, i, S}^{(I, h)} RDS_{t}^{(I, h)}  \nonumber \\
& + & \sum_{h \in\{\tr, \p\}} \beta_{t, i, K}^{(I, h)} RDK_{t}^{(I, h)}.
\end{eqnarray}
The model labeled as \textbf{6} combines the market and average idiosyncratic heterogeneously persistent higher moments, and controls for the variables from the FF3 model
\begin{eqnarray}
r_{t, i} & =  & \sum_{r \in \{ m, I \}} \sum_{h \in\{\tr, \p\}} \beta_{t, i, V}^{(r, h)} RDVOL_{t}^{(r, h)} +  \sum_{r \in \{ m, I \}} \sum_{h \in\{\tr, \p\}} \beta_{t, i, S}^{(r, h)} RDS_{t}^{(r, h)}  \nonumber \\
& + & \sum_{r \in \{ m, I \}} \sum_{h \in\{\tr, \p\}} \beta_{t, i, K}^{(r, h)} RDK_{t}^{(r, h)} + \sum_{j = 1}^J \gamma_{t, i, j} Z_{t, i, j}.
\end{eqnarray}

\newpage
\section{Additional figures and tables}
\label{sec:add_fig}

\begin{table}[p!]
\footnotesize
\centering
\caption[]{The cross-section of stocks - robustness checks}
\begin{minipage}{\textwidth} 
\footnotesize
We report the lambda coefficients from the second-stage regression specified by Equation~(\ref{eq:model_stage2}). The prices of risk specified in Equation~(\ref{eq:model_stage1}) are estimated using daily stocks data. The full list of model specifications is provided in Appendix~\ref{app:models}. We use 5 USD threshold in the data cleaning procedure. The t-statistics displayed in parentheses are computed using Newey-West standard errors.
\end{minipage}
\vspace{\medskipamount}

\begin{tabular}{l c c c c c c }
\hline
\hline
\textbf{Variable} & \textbf{1} & \textbf{2} & \textbf{3} & \textbf{4} & \textbf{5} & \textbf{6} \\
\hline
const & 0.000217 & 0.00024 & 0.000296 & 0.000245 & 0.000264 & 0.000302 \\  & (2.008895) & (2.327188) & (3.373519) & (2.333033) & (2.653727) & (3.373021) \\
$RVOL_{t}^{(m)}$ & -0.000101 &  & -9.7e-05 &  &  &  \\  & (-1.357097) &  & (-1.523763) &  &  &  \\
$RS_{t}^{(m)}$ & -0.007933 &  & 0.00025 &  &  &  \\  & (-0.777464) &  & (0.037301) &  &  &  \\
$RK_{t}^{(m)}$ & 0.001721 &  & 0.013596 &  &  &  \\  & (0.069695) &  & (0.681716) &  &  &  \\
$RVOL_{t}^{(m, \tr)}$ &  &  &  & -0.000134 &  & -0.000105 \\  &  &  &  & (-1.868343) &  & (-1.794311) \\
$RVOL_{t}^{(m, \p)}$ &  &  &  & 1.7e-05 &  & 9e-06 \\  &  &  &  & (0.773166) &  & (0.518574) \\
$RS_{t}^{(m, \tr)}$ &  &  &  & -0.00556 &  & -0.000606 \\  &  &  &  & (-0.590639) &  & (-0.09459) \\
$RS_{t}^{(m, \p)}$ &  &  &  & 0.000233 &  & 4.7e-05 \\  &  &  &  & (0.762307) &  & (0.172978) \\
$RK_{t}^{(m, \tr)}$ &  &  &  & -0.00598 &  & 0.01072 \\  &  &  &  & (-0.272689) &  & (0.590803) \\
$RK_{t}^{(m, \p)}$ &  &  &  & -0.001368 &  & -0.001982 \\  &  &  &  & (-0.557476) &  & (-0.990648) \\
$RVOL_{t}^{(I)}$ &  & -0.000182 & -0.000202 &  &  &  \\  &  & (-1.747955) & (-2.0383) &  &  &  \\
$RS_{t}^{(I)}$ &  & -0.003078 & -0.001073 &  &  &  \\  &  & (-0.951594) & (-0.443949) &  &  &  \\
$RK_{t}^{(I)}$ &  & -0.005373 & -0.000628 &  &  &  \\  &  & (-0.881274) & (-0.126726) &  &  &  \\
$RVOL_{t}^{(I, \tr)}$ &  &  &  &  & -0.000198 & -0.000183 \\  &  &  &  &  & (-2.070916) & (-1.994129) \\
$RVOL_{t}^{(I, \p)}$ &  &  &  &  & -4e-06 & -1.9e-05 \\  &  &  &  &  & (-0.094453) & (-0.578467) \\
$RS_{t}^{(I, \tr)}$ &  &  &  &  & -0.003104 & -0.001315 \\  &  &  &  &  & (-1.00532) & (-0.564864) \\
$RS_{t}^{(I, \p)}$ &  &  &  &  & -0.000139 & -9.9e-05 \\  &  &  &  &  & (-1.506872) & (-1.189938) \\
$RK_{t}^{(I, \tr)}$ &  &  &  &  & -0.005456 & -0.000735 \\  &  &  &  &  & (-0.986052) & (-0.158903) \\
$RK_{t}^{(I, \p)}$ &  &  &  &  & -0.000837 & -0.00025 \\  &  &  &  &  & (-0.72548) & (-0.236493) \\
\hline
\end{tabular}

\label{tab:crsp6m5usd}
\end{table}

\end{document}